\documentclass[twocolumn,amsmath,amssymb,floatfix,superscriptaddress,prb,longbibliography]{revtex4-2}
\usepackage{hyperref}
\usepackage{bm}
\usepackage{amsmath}
\usepackage{booktabs}

\usepackage{graphicx}
\usepackage{xcolor}
\usepackage{siunitx}
\usepackage[version=4]{mhchem}
\usepackage{ulem}

\begin{document}

\preprint{APS/123-QED}

\title{Enhancement of local magnetic ordering temperature due to non-magnetic disorder in anti-Heusler alloy Al$_2$MnCu}
\author{Soumya Bhowmik}
\email{soumyabhowmikwb@gmail.com}
\affiliation{Condensed Matter Physics Division, Saha Institute of Nuclear Physics,
              A CI of Homi Bhabha National Institute, 1/AF Bidhannagar, Kolkata 700064, India}
\author{Suvayan Saha}
\affiliation{Condensed Matter Physics Division, Saha Institute of Nuclear Physics,
              A CI of Homi Bhabha National Institute, 1/AF Bidhannagar, Kolkata 700064, India}
\author{Saikat Debnath}
\affiliation{Department of Physics, M.V. College, Buxar, Bihar 802101, India}
\author{Arka Patra}
\affiliation{Condensed Matter Physics Division, Saha Institute of Nuclear Physics,
              A CI of Homi Bhabha National Institute, 1/AF Bidhannagar, Kolkata 700064, India}
\author{Biswarup Satpati}
\affiliation{Condensed Matter Physics Division, Saha Institute of Nuclear Physics,
              A CI of Homi Bhabha National Institute, 1/AF Bidhannagar, Kolkata 700064, India}
\author{Maxim Avdeev}
\affiliation{Australian Nuclear Science and Technology Organisation, Locked Bag 2001, Kirrawee DC, New South Wales 2232, Australia}
\affiliation{School of Chemistry, The University of Sydney, Sydney, New South Wales 2006, Australia}
\author{Shishir Kumar Pandey}
\email{shishir.kr.pandey@gmail.com}
\affiliation{Department of General Sciences (Physics), Birla Institute of Technology and Science, Pilani, Dubai Campus, Dubai International Academic City, Dubai 345055, United Arab Emirates}
\affiliation{Department of Physics, Birla Institute of Technology and Science, Pilani, Hyderabad Campus, Jawahar Nagar, Kapra Mandal, Medchal District, Telangana 500078, India}
\author{Chandan Mazumdar}
\email{chandan.mazumdar@saha.ac.in}
\affiliation{Condensed Matter Physics Division, Saha Institute of Nuclear Physics,
              A CI of Homi Bhabha National Institute, 1/AF Bidhannagar, Kolkata 700064, India}

\date{\today}

\begin{abstract}
Structural disorder is generally regarded as detrimental to magnetism. Since magnetic properties are governed by both the arrangement of magnetic atoms and the crystal environment, including non-magnetic atoms that mediate exchange interactions, the distinct role of magnetic and non-magnetic disorder remains less understood. The recently reported anti-Heusler alloy, Al$_2$MnCu, forming with $\sim$50\% intra-symmetry site disorder (B2) involving the magnetic Mn atom and $\sim$5\% inter-symmetry site disorder associated with non-magnetic Al and Cu atoms, offers a good platform to disentangle the individual roles of magnetic and non-magnetic atomic disorder. Our magnetic critical analysis, in the vicinity of the Curie temperature ($T_{\rm C}$), does not lead to any conventional single universality class expected for a standard ferromagnet; instead, the non-monotonic temperature dependence of effective critical exponents points towards more complex magnetic interactions, akin to inhomogeneous ferromagnets. Dc magnetometry complemented by higher-harmonic ac magnetic susceptibility data, further confirms the persistence of locally ordered magnetic clusters above $T_{\rm C}$. Theoretical analysis establishes that the enhancement of the local magnetic ordering temperature arises directly from the strengthening of exchange interactions due to the small degree of non-magnetic disorder, whereas $T_{\rm C}$ is hardly influenced by the much larger level of magnetic-atom disorder. The combined experimental and theoretical investigations on the anti-Heusler alloy Al$_2$MnCu reveal and explain the unusual observation of nonmagnetic disorder-induced elevation of $T_{\rm C}$, providing a paradigm for the counterintuitive role of disorder in magnetic materials.

\end{abstract}
\maketitle

\section{Introduction}
The phenomenon of structural disorder is inevitable in material science. These disorders, such as lattice strain, chemical impurities, vacancies, dopants, or antisite swap disorder, etc., can have a high impact on magnetic~\cite{PhysRevB.106.134423, PhysRevB.96.144423, PhysRevB.108.054430}, transport~\cite{Garmroudi2022}, and optical~\cite{PhysRevLett.47.1480} properties of materials. In magnetism, for example, the positive influence of disorder can be found in some cases, e.g., where altermagnetism is favoured over antiferromagnetism through compressive strain~\cite{León2025}, chemical impurities stabilise the skyrmion phase in a wider temperature range~\cite{PhysRevB.102.134410}, atomic disorder-induced mimicry of quantum spin liquid state~\cite{PhysRevLett.119.157201} etc. However, despite those few positive impacts, structural disorder more commonly hinders primary exchange pathways, restricts magnetic correlation lengths, and consequently disrupts the long-range magnetic order. It reduces the magnetic moment~\cite{PhysRevB.97.165137}, decreases the ordering temperature~\cite{PhysRevB.77.195104}, or even completely diminishes the long-range interaction~\cite{PhysRevB.107.184408}. Hence, in the current era of quantum materials, considerable efforts are being devoted to enhancing the ordering temperature of spin-dependent phenomena, such as superconductivity and spin-gapless semiconductivity, toward room temperature. In this context, understanding how intrinsic structural disorder influences the magnetic ordering temperature is important from both fundamental and technological perspectives.

In the study of disorder-dependent phenomena, Heusler alloys ($X_2YZ$, where $X$ and $Y$ are the $d$- and $Z$ is the $p$-block element) are a promising class of materials as they are intrinsically susceptible to antisite disorder owing to the similar atomic radii and electronic structures of their constituent elements~\cite{PhysRevB.108.245151}. These compounds manifest many intriguing phenomena, such as Weyl semimetallicity~\cite{doi:10.1073/pnas.1810842115}, half-metallic ferromagnetism~\cite{PhysRevB.108.045137, PhysRevB.106.115148}, spin-gapless semiconductivity~\cite{PhysRevLett.110.100401}, spin glass~\cite{PhysRevB.107.184408, PhysRevB.108.054405}, etc., which are reported to be highly sensitive to antisite disorder. Hence, many of these compounds have already been investigated in detail for the influence of atomic disorder on their physical properties~\cite{GRAF20111}.

Recently, a new class of Heusler compounds, named anti-Heusler alloys with a reversed formula unit $Z_2XY$, has emerged~\cite{PhysRevB.111.174417}. To date, very few anti-Heusler compounds have been reported, and almost all of them exhibit re-entrant spin-glass behaviour~\cite{PhysRevB.97.184421, Samanta2022}. In the anti-Heusler Al$_2$MnFe, both inter- and intra-symmetry site disorder involve the magnetic atom Mn, with the former being particularly responsible for the emergence of the low-temperature spin-glass phase~\cite{463q-ph2y}. Conversely, Al$_2$MnCu possesses these two different types of disorder with fundamentally different characteristics. The intra-symmetry site disorder (50\%) involves the magnetic atom Mn along with Cu, whereas the inter-symmetry site disorder (5\%) is not associated with any magnetic atom, but rather with non-magnetic Cu and Al atoms~\cite{PhysRevB.111.174417}. In general, the magnetic properties of a material depend either on the arrangement of magnetic atoms themselves or on the crystal environment, comprising other non-magnetic atoms, which can mediate the interactions. Thus, the intrinsic coexistence of two distinct types of disorder in Al$_2$MnCu, involving magnetic and non-magnetic atoms, provides an ideal platform for investigating the respective roles of magnetic and non-magnetic disorder within a single compound.

Our detailed investigation of the magnetic properties of Al$_2$MnCu suggests the compound to be an inhomogeneous ferromagnet with complex interactions, as the critical exponents obtained from magnetic critical analysis do not follow any conventional universality class. DC magnetisation, complemented by AC higher-harmonic magnetic susceptibility, further reveals that, even above the bulk Curie temperature, magnetic clusters having stronger interaction strength and higher local ordering temperature persist in the midst of a global paramagnetic matrix. This phenomenon is known as Griffiths phase behaviour. Theoretical calculations demonstrate that the enhancement of local magnetic ordering temperature is a direct consequence of inter-symmetry site disorder involving non-magnetic atoms, whereas a substantially higher degree of B2-type disorder involving magnetic atoms does not influence the Curie temperature at all. Structural disorder is generally considered detrimental to long-range magnetic order and is consequently expected to suppress the magnetic transition temperature ($T_{\rm C}$). In contrast, our results demonstrate a counterintuitive disorder-induced enhancement of the local ordering temperature, highlighting a rare case in which structural disorder can promote rather than suppress magnetic ordering.

\begin{figure*}[ht]
  \centering
  \includegraphics[width=\textwidth]{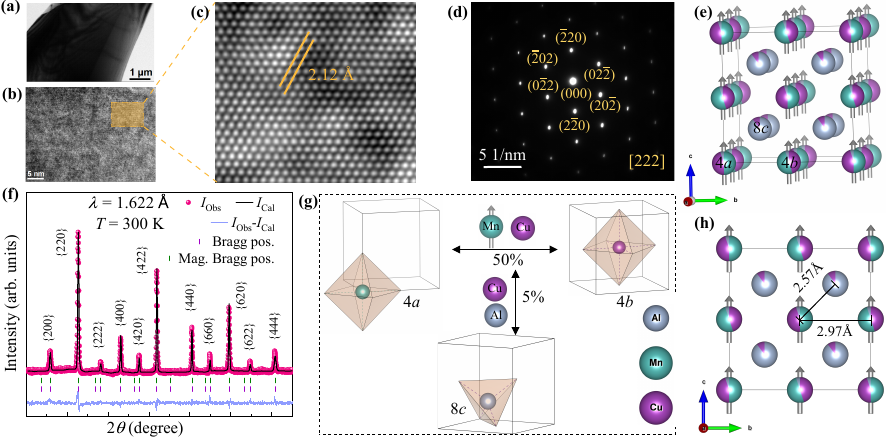}
  \caption{(a) and (b) High-resolution transmission electron microscope image with a zoomed version (c). (d) Selected area electron diffraction pattern. (e) Crystal structure obtained from Rietveld refinement of the room-temperature neutron diffraction spectra (f). (g) Schematic image of different symmetry sites and the atom-swapping disorder between them. (h) Crystal structure viewed along the $a$-axis.}
  \label{Fig:structure}
  \end{figure*}

\section{Methods}
\subsection{Experimental details}
The polycrystalline anti-Heusler compound Al$_2$MnCu is synthesized by utilizing a tri-arc furnace. The constituent elements Al, Mn and Cu, having purity $>$99.9 wt.\% ( M/s Alfa Aesar, USA), in a molar ratio of 2:1:1, were melted in an atmosphere with continuous Argon gas flow. To obtain a homogeneous compound, the ingot was remelted 5-6 times, with the ingot flipped after each melting. The neutron diffraction (ND) measurement was carried out at the ECHIDNA beamline with a wavelength, $\lambda$ = 1.622 \AA \, at ANSTO, Australia, at $T$= 300~K~\cite{Avdeev:ks5608}, and the spectra have been analysed by Full-Prof software package~\cite{RODRIGUEZCARVAJAL199355} . For Transmission electron microscopy (TEM) measurement, the sample is prepared following the standard protocol of mechanical thinning, dimpling and final thinning utilizing a precision-ion-polishing system (PIPS, Gatan, Pleasanton, CA). The ion-polish was done at an energy of 3.0 keV and final cleaning was made at 1.2 keV energy. The high-resolution transmission electron microscopy (HRTEM) was carried out employing an FEI Tecnai G2 F30, S-Twin microscope, which operated at 300 kV and was attached to a GATAN Orius SC1000B CCD camera. The dc magnetization property was measured in a physical property measurement system (M/s Quantum Design, Inc., USA) with a field range of 0$-$70~kOe in the temperature regime 5$-$400~K utilizing the standard zero-field-cooled (ZFC) and field-cooled (FC) protocols. In the ZFC process, the compound is cooled down to its desired temperature from far above its paramagnetic region with zero applied field, and the magnetic measurements start after applying a magnetic field, whereas in the FC protocol, the cooling process was carried out by applying the same constant magnetic field, which will be utilized during measurements. In isothermal magnetization, the applied magnetic field has been corrected for the demagnetization factor. The linear ac susceptibility data have been acquired in the temperature range 275$-$385~K, with an applied dc field ($H_{dc}$) of 0$-$300~Oe and a driving ac field amplitude ($h_{ac}$) of 1.5$-$13~Oe, while the frequency range was 113$-$9951~Hz in the same PPMS with an ACMS attachment.

\subsection{Computational Methodology}

\subsubsection{First-Principles Calculations}
First-principles density functional theory (DFT) calculations were carried out using the projector augmented-wave (PAW) method \cite{PAW, PAWpotentials1}, as implemented in the Vienna \textit{ab initio} Simulation Package (VASP) \cite{Kresse}. The generalized gradient approximation (GGA) parameterized by Perdew, Burke, and Ernzerhof (PBE) \cite{PBE} was utilized to describe the exchange-correlation interactions. The PAW pseudopotentials explicitly treated 3 valence electrons for Al ($3s^2 3p^1$), 11 valence electrons for Cu ($3d^{10} 4s^1$), and 13 valence electrons for Mn ($3p^6 3d^5 4s^2$).

The Kohn-Sham wave functions were expanded using a plane-wave basis set with an energy cutoff of 400 eV. To ensure a high degree of numerical accuracy, the energy convergence criterion for the self-consistent electronic loops was strictly set to $10^{-5}$~eV. Brillouin zone integration was performed using a  Monkhorst-Pack $k$-point mesh. For the pristine unit cell and the standard cubic B2-disordered cell, a dense $15 \times 15 \times 15$ $k$-point grid was used. For the larger $1 \times 1 \times 2$ supercells employed to simulate specific atomic mixing (e.g., 12.5\% disorder), the mesh was appropriately scaled to $15 \times 15 \times 7$ to maintain a consistent $k$-point sampling density across the different geometries. A Gaussian smearing method with a width of 0.05~eV was employed to determine partial occupancies near the Fermi level. The structures considered for three different phases, namely, L2$_1$ ordered, B2-type disordered and B2+12.5\% Cu-Al disorder, in our Spin-polarized calculations are shown in Fig. S7 of the Supplementary Materials (SM) ~\cite{SM}.

\subsubsection{Magnetic Exchange Interactions}
To evaluate the magnetic exchange couplings ($J_{ij}$) between the localised magnetic moments, we employed a multiscale approach combining Wannier function-based tight-binding (TB) model \cite{wannier90} and the magnetic force theorem. The spin-polarised $ab$ $initio$ band structure was projected onto a TB Hamiltonian with basis formed by $s$ and $p$ orbitals of Al; the $s$, $p$, and $d$ orbitals of Cu; and the $d$ orbitals of Mn.

Using this TB model, the Heisenberg exchange parameters were calculated via the Green's function method based on the magnetic force theorem, utilizing the TB2J Python package \cite{tb2j}. In these calculations, the non-magnetic orbitals associated with the Al and Cu species were systematically projected out to strictly isolate the effective $d$-$d$ exchange couplings mediating the interactions between the Mn atoms. The exchange parameters were computed considering interacting pairs up to a real-space cutoff distance of 11.0~\AA.

\subsubsection{Atomistic Spin Dynamics}
Macroscopic magnetic properties, mainly the Curie temperature ($T_C$), were estimated by applying atomistic spin dynamics simulations using the VAMPIRE package \cite{vampire}. The simulations were carried out on a large three-dimensional computational cell of $15 \times 15 \times 15$~nm$^3$ initialized with periodic boundary conditions. The temperature-dependent magnetization plot was generated by sweeping the simulated temperature from 0~K to 1000~K in steady increments of 25~K, considering the Heisenberg model $H = -\frac{1}{2}\sum_{i\ne j}J_{ij} \Vec{S_i}\cdot \Vec{S_j}$. At each temperature step, the spin configuration was allowed to equilibrate for 2,500 time steps, followed by an additional 3,000 loop time steps to gather the time-averaged macroscopic magnetization and the Curie temperature.

\section{Experimental Results}
\subsection{Structural details}

To examine the crystallinity of the synthesized polycrystalline compound,  high-resolution transmission electron microscopy (HRTEM) measurements have been carried out. The resultant HRTEM images are shown in Fig~\ref{Fig:structure}(a) and Fig~\ref{Fig:structure}(b), with a zoomed-in picture of the (220) lattice plane, where the lattice spacing is estimated to be 2.12 \AA\,[Fig~\ref{Fig:structure}(c)]. A selected area electron diffraction (SAED) pattern from an arbitrary area is also shown in Fig~\ref{Fig:structure}(d), confirming the single-crystalline nature of the material. SAED was indexed using the lattice parameters of the cubic crystal structure, indicating that the lamella was oriented along the [222] direction.

A complementary analysis of crystal structure is neutron diffraction, which reveals the atomic arrangements in a crystal's unit cell. The room-temperature neutron diffraction spectra of Al$_2$MnCu have been analyzed by cubic structure with lattice parameters $a$= 5.947(7) \AA, and space group ${Fm\bar{3}m}$ [Fig~\ref{Fig:structure}(f)]. The nuclear structure is obtained as 50\% Mn–Cu disorder (B2-type), together with an additional $\sim$5\% Cu–Al atomic disorder, while the magnetic structure is ferromagnetic, with a moment of 0.93 $\mu_B$/f.u. at $T = 300$~K, associated only with the Mn atoms, consistent with the earlier report~\cite{PhysRevB.111.174417}. This atomic arrangement, i.e., B2+5\% Cu-Al atom swapping, implies essentially two different types of disorders: a $\sim$50\% intra-symmetry site disorder (B2-type), constrained in octahedral symmetry sites only (Wyckoff position: 4$a$$-$4$b$) and another one with $\sim$5\% inter-symmetry site disorder between octa- \& tetra-hedral symmetry sites (Wyckoff position: 4$a/4b$$-$8$c$) [Fig~\ref{Fig:structure}(g)]. The B2-type intra-symmetry site disorder involves the magnetic atom Mn along with Cu, while the inter-symmetry site disorder is associated with the non-magnetic Cu and Al atoms. Detailed atomic occupancies are tabulated in Table S1 of the SM ~\cite{SM}. A close scrutiny of the crystal structure [Fig~\ref{Fig:structure}(e)], particularly its projection on (100) plane [Fig~\ref{Fig:structure}(h)], reveals that the magnetic Mn atoms (at 4$a$ or 4$b$ positions, due to B2 disorder) are surrounded by eight non-magnetic Cu/Al atoms occupying 8$c$ positions at a distance of 2.57 \AA. In contrast, for the same central Mn atoms, other Mn atoms can be found as their second-nearest neighbour at a slightly larger distance of 2.97 \AA. The closer placement of nonmagnetic Cu/Al atoms to magnetic Mn atoms indicates that the disorder at the 8$c$ site would likely influence the magnetic exchange interaction quite significantly.

\subsection{Magnetic properties}
\subsubsection{dc magnetization}
\label{dc magnetization}
\begin{figure*}[ht]
  \includegraphics[width=\textwidth]{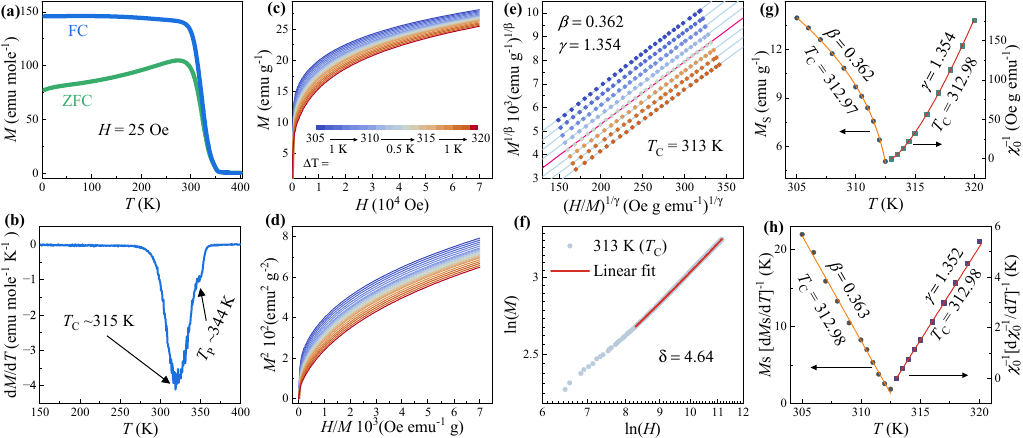}
    \caption{(a) Temperature dependent magnetization, $M(T)$, data measured at 25~Oe applied field in ZFC and FC protocol. (b) d$M$/d$T$ curve of FC data. (c) Field-dependent magnetization, $M(H)$, data at various temperatures ($305\leq T \leq 320~K$). (d) Arrott plot of the isotherms around $T_{\rm C}$. (e) Modified Arrott plot of the isotherms. Blue lines are linear fits, and the red line corresponds to $T_{\rm C}$. For clarity, only a few curves are shown. (f) Critical isotherm, $M(H)$, represented on a log$_e$-log$_e$ scale, and the solid line is the linear fit. (g)  Temperature variation of $M_{\rm S}$ (left axis) and $\chi_{\rm 0}^{-1}$ (right axis), and the fits (solid lines) according to Eqs.~\ref{eq.Ms} and ~\ref{eq.Chi-1}. (h) Kouvel-Fisher plot and corresponding fit following Eqs.~\ref{KF1} and ~\ref{KF2}.}
    \label{Fig:Critical}
  \end{figure*}

Temperature-dependent dc magnetization, $M(T)$, measured in zero-field-cooled (ZFC) and field-cooled (FC) protocols in the temperature range 5-400~K, under an applied field $H$= 25~Oe, is presented in Fig~\ref{Fig:Critical}(a). The $M(T)$ curve suggests a long-range ferromagnetic (FM) nature, where the Curie temperature, $T_{\rm C}$, is determined from the dip of the d$M$/d$T$ curve as $\sim$315~K [Fig~\ref{Fig:Critical}(b)], consistent with the previous report~\cite{PhysRevB.111.174417}. On close inspection, an additional anomaly is observed in the d$M$/d$T$ curve around 344~K, slightly above the bulk Curie temperature of this material. As the strength of the externally applied magnetic field gradually increases, this anomaly also gets suppressed, and for $H>$ 500~Oe, the signature vanishes completely [Fig.~S5 of SM~\cite{SM}]. Such an observation restricted to the very low magnetic field region suggests that the origin of this anomaly is magnetically short-range in nature, which will be addressed in detail later.

To gain a deeper understanding of FM interactions, a magnetic critical scaling analysis is carried out. In a ferromagnet-to-paramagnetic phase transition, critical analysis near $T_{\rm C}$ offers deep insight into spin dimensionality, correlation length, magnetic interactions, etc. To investigate the critical behavior of Al$_2$MnCu, isothermal magnetization, $M(H)$, data were acquired in the temperature range 305 $\leq T \leq$ 320~K with a temperature interval $\triangle T$=1~K. For a better analysis, in the vicinity of $T_{\rm C}$, the data were taken at a much closer temperature interval, $\triangle T$=0.5~K [Fig~\ref{Fig:Critical}(c)]. According to the Landau theory of FM-PM phase transition, the Gibbs free energy is $G(T, M)= G_0+ \frac{a}{2}M^2+ \frac{b}{4}M^4 - MH$, where $a$ and $b$ are temperature-dependent parameters~\cite{PhysRevB.105.024419}. In equilibrium state ${\partial G}/{\partial M}= 0$, the magnetic equation of state is written as
\begin{equation}
    H/M= a+bM^2
\end{equation}
Subsequently, all the isotherms are plotted in the $M^2$ vs $H/M$ representation [Fig.~\ref{Fig:Critical}(d)], called the Arrott plot~\cite{ArrottPlot, Patra_2026}. According to Banerjee's criterion, the positive (negative) slope of the Arrott plots implies a second (first)- order phase transition~\cite{BANERJEE196416}. In Al$_2$MnCu, the positive slope throughout the critical region in the Arrott plot indicates that the phase transition is second-order. Further, for the correct choice of theoretical model in describing the phase transition, the plot should appear as a parallel set of straight lines where the critical isotherm (at $T_{\rm C}$) passes through the origin~\cite{PhysRevB.105.214407}. However, the nonlinear nature of the isotherms, even at high magnetic fields up to 70 kOe, failed to describe the critical nature of the compound through Arrott plots and Landau mean-field theory. Therefore, a modified Arrott plot (MAP) method
\begin{equation}
  (H/M)^{1/\gamma}=a \left(\frac{T-T_{\rm C}}{T_{\rm C}}\right)+bM^{1/\beta}
\end{equation}
where $\beta$, $\gamma$, and $\delta$ are the critical exponents, has been implemented~\cite{PhysRevLett.19.786}. In the critical regime of a second-order phase transition, the correlation length $\xi= \xi_0\left| \left (T-T_{\rm C}\right)/T_{\rm C}\right|^{-\nu}$ diverges~\cite{PhysRevB.109.054413}. According to the universal scaling law, spontaneous magnetization $M_{\rm S}$ below $T_{\rm C}$, inverse initial susceptibility $\chi_{\rm 0}^{-1}$ above $T_{\rm C}$ and magnetization, $M$ at $T_{\rm C}$ exhibits following power-law dependence:
\begin{eqnarray}
M_{\rm S}(T) \propto (-\epsilon)^\beta,\:\epsilon<0, \label{eq.Ms}\\
\chi_{\rm 0}^{-1}(T) \propto \epsilon^{\gamma},\:\epsilon>0, \label{eq.Chi-1}\\
M(H, T_{\rm C}) \propto H^{1/\delta},\:\epsilon = 0 \label{eq.Delta}
\end{eqnarray}

\noindent where $\epsilon = (\frac{T-T_{\rm C}}{T_{\rm C}})$ is reduced temperature~\cite{stanley1971introduction}. The critical exponents are obtained by simultaneously satisfying the above power-law equations and the MAP with an iterative approach. After examining all the universality classes, the initial parameters have been chosen as $\beta$= 0.365 and $\gamma$= 1.386 (3D Heisenberg model), for which the most parallel quasi-straight lines were formed [details are discussed in section S1 of the SM~\cite{SM}].  With these initial parameters obtained from MAP, the critical exponent values have been modified through a power-law equation; these modified exponents are utilised in the next iteration in reconstructing MAP, until the values converge~\cite{PhysRevB.105.214407}. After subsequent iterations, the yielded critical exponent values become $\beta=0.362$(3) and $\gamma=1.354(4)$, which formulates the MAP as a set of parallel straight lines, where the critical isotherm passes through the origin, and the corresponding temperature is considered to be $T_{\rm C}$ = 313~K [Fig.~\ref{Fig:Critical}~e and Fig.~S3 of SM~\cite{SM}] that simultaneously satisfies the power-law equations [Fig.~\ref{Fig:Critical}(g)]. Furthermore, the $\beta$ and $\gamma$ values, obtained from the above iteration, are verified through Kouvel-Fisher (KF) relations~\cite{KFPlot}:
\begin{eqnarray}
    M_{\rm S}(dM_{\rm S}/dT)^{-1}= (T_{\rm C}-T)/\beta \label{KF1}\\
    \chi_{\rm 0}^{-1}(d\chi_{\rm 0}^{-1}/dT)^{-1}=(T-T_{\rm C})/\gamma \label{KF2}
\end{eqnarray}
Accordingly, temperature dependence $M_{\rm S}(dM_{\rm S}/dT)^{-1}$ and $\chi_{\rm 0}^{-1}(d\chi_{\rm 0}^{-1}/dT)^{-1}$ have been plotted having slopes are 1/$\beta$ and 1/$\gamma$ respectively, and the intercepts yield $T_{\rm C}$ [Fig.~\ref{Fig:Critical}(h)]. The determined $\beta$= 0.363(6), $\gamma$= 1.352(2), and $T_{\rm C}$= 312.98(6) values substantiate the critical exponent parameters obtained from the MAP and power-law equations (Table~\ref{tab:critical}). Another critical exponent $\delta$= 4.64(4) is estimated from the slope (1/$\delta$) of the linear fit in the high-field region of the $M(H)$ curve at $T_{\rm C}$ on a log$_{10}$-log$_{10}$ scale [Fig.~\ref{Fig:Critical}(f)]. Moreover, using the $\beta$ and $\gamma$ obtained from MAP and KF method, $\delta$ is also calculated by Widom scaling relation~\cite{10.1063/1.1696618}, $\delta=1+\gamma/\beta$, to be 4.75(3) and 4.69(6), respectively. These are close to the value determined from the critical isotherm, further supporting that the critical exponents and $T_{\rm C}$ are self-consistent and reliable.

\begin{table*}
\centering
\begin{tabular}{lccccccc}
\hline\hline
Model/Material & $\{d:n\}$ & Technique & $\beta$ & $\gamma$ & $\delta$ & $T_{\rm C}$ (K) & Ref. \\
\hline
Mean field & & Theory & 0.5 & 1 & 3 & & \cite{PhysRevB.32.7209} \\
3D Heisenberg & $\{3:3\}$ & Theory & 0.365 & 1.386 & 4.8 & & \cite{PhysRevB.65.144520} \\
3D XY & $\{3:2\}$ & Theory & 0.345 & 1.316 & 4.81 & & \cite{PhysRevB.21.3976} \\
3D Ising & $\{3:1\}$ & Theory & 0.325 & 1.241 & 4.82 & & \cite{BANERJEE196416} \\
2D Ising & $\{2:1\}$ & Theory & 0.125 & 1.75 & 15 & & \cite{PhysRevB.65.064443} \\
Tricritical mean field & & Theory & 0.25 & 1 & 5 & & \cite{KAUL19855} \\
Al$_2$MnCu & & MAP & 0.362(3) & 1.354(4) & 4.75(3) & 313 & This work \\
& & KF & 0.363(6) & 1.352(2) & 4.69(6) & 312.98(6) & \\
& & CI & & & 4.64(4) & & \\
\hline\hline
\end{tabular}
\caption{Critical exponents obtained for Al$_2$MnCu compared with theoretical values.}
\label{tab:critical}
\end{table*}

\begin{figure*}[ht]
  \centering
  \includegraphics[width=\textwidth]{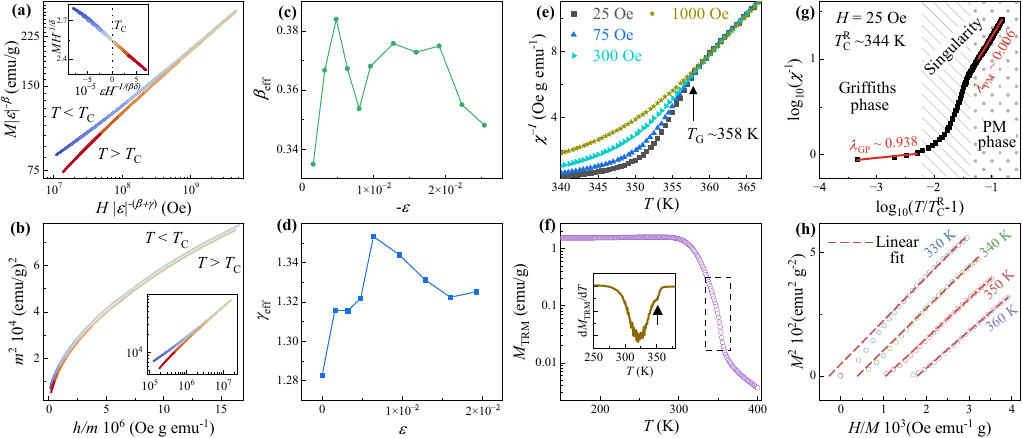}
  \caption{(a) Scaling plot of $M|\epsilon|^{-\beta}$ vs. $H|\epsilon|^{-(\beta+\gamma)}$ in log$_{10}$-log$_{10}$ scale. The inset is a modified scaling plot of $MH^{-1/\delta}$ vs  $\epsilon H^{-1/(\beta \delta)}$ (b) Scaling plot in $m^2$ vs. $h/m$ form. The inset is the same data in log$_{10}$-log$_{10}$ scale representation. Effective critical exponent, (c) $\beta_{eff}$  and (d) $\gamma_{eff}$  are plotted with the function of reduced temperature, $\epsilon$. (e) Temperature-dependent inverse susceptibility, $\chi^{-1}(T)$ for various applied field. (f) Thermoremanent magnetization curve and inset show its temperature derivative. (g) Double-logarithmic plot of inverse susceptibility data at 25~Oe with respect to reduced temperature. The solid lines are fit with the power-law equation, described in the text. (h) Arrott plot in the temperature region of $T_{\rm C} \leq T \leq T_{\rm G}$.
  }
  \label{Fig:Griffiths Phase}
  \end{figure*}

The trueness of the generated exponents must be examined by scaling analysis. In the asymptotic critical region, the scaling relation is written as~\cite{stanley1971introduction}:
\begin{equation}
    M(H, \epsilon)= \left| \epsilon \right|^\beta f_{\pm}\left[\frac{H}{|\epsilon|^{(\gamma+\beta)}}\right]
\end{equation}
where $f_+$ ($f_-$) is a regular function for $T>T_{\rm C}$ ($T<T_{\rm C}$). This scaling equation can be reexpressed as $m=f_{\pm}(h)$ with renormalized magnetization $m~(=M|\epsilon|^{-\beta}$) and renormalized field $h~(=H|\epsilon|^{-(\beta+\gamma)})$. The scaling equation suggests that with a proper value of the critical exponent, the $m$ vs. $h$ plot collapses into two separate universal curves; one for $T>T_{\rm C}$ and another for $T<T_{\rm C}$. To achieve such scaling behaviour, the critical exponents must be true and unique. The $m$ vs. $h$ plot, presented in Fig.~\ref{Fig:Griffiths Phase}(a), depicts that all the data indeed fall into two branches, implying the accuracy of the obtained values of $\beta$, $\gamma$, and $\delta$. The values of the exponents and $T_{\rm C}$ are further verified by plotting $m^2$ vs. $h/m$, where too the isotherms collapse onto two separate branches [Fig.~\ref{Fig:Griffiths Phase}(b)]. Besides, another form of scaling equation: $\frac{M}{H^\delta}=k(\frac{\epsilon}{H^{1/\beta}})$ has also been applied~\cite{PhysRevB.109.094408} to obtain a $MH^{-1/\delta}$ vs $\epsilon H^{-1/\beta \delta}$ plot [Fig.~\ref{Fig:Griffiths Phase}(a)~(inset)], where all isotherms fall onto a single universal curve. Together, all these scaling analyses unequivocally substantiated the reliability and accuracy of the estimated critical exponents.


These critical exponents, established through the aforementioned rigorous analysis, reveal that the compound does not follow any conventional universality; rather, $\beta$ is smaller/close to the 3D Heisenberg~\cite{PhysRevB.65.144520}, whereas $\gamma$ is slightly higher/close to the 3D XY model~\cite{PhysRevB.21.3976}. Overall, the exponents lie between the 3D Heisenberg and 3D XY models. One of the possibilities for the deviation from the known set of universal classes is that the critical temperature region chosen for analysis is far from the critical temperature~\cite{PhysRevB.99.214414}. To check this proposition, the validity of the temperature region that has been employed during the analysis needs to be examined. Here, the critical analysis is carried out in the reduced temperature region of $3.4 \times 10^{-3}\leq \lvert \epsilon \rvert \leq 2.2 \times 10^{-2}$, which falls well inside the asymptotic critical regime of $\lvert \epsilon \rvert \simeq 10^{-2}$ and hence the departure from universal classes can not be associated with the chosen temperature regime~\cite{PhysRevB.51.12585}. Another possibility is that the critical exponents can be shifted towards the mean-field value below a certain value of reduced temperature, $\epsilon_{\rm {co}}$, due to the isotropic long-range dipolar interactions~\cite{PhysRevLett.30.559}. The crossover value, $\epsilon_{\rm {co}}$, depends on the strength of dipolar interactions. The $\epsilon_{\rm {co}}$ is determined as $\epsilon_{\rm {co}}\approx g_d^{1/\phi_d}$, where $g_d$ and $\phi_d$ represent the dipolar interaction strength and the crossover exponent, respectively~\cite{PhysRevB.8.3323}. Here, $\phi_d \approx$ Heisenberg exponent, $\gamma_H = 1.386$ and $g_d$ can be estimated as $g_d \approx \frac{0.87}{T_{\rm C}} \frac{p^2\theta}{v}$, where $p=g \sqrt{S(S+1)}$, $\theta$ is the correction factor, and $v$ is the unit-cell volume~\cite{RIED1992259}. The studied compound Al$_2$MnCu has a unit-cell volume, $v$ = 210.32 $\AA^3$ and the moment-bearing ion is Mn$^{+3}$ ($S$ = 2); hence, $p$ = 4.899 $\mu_B$. For S =2, $\theta$ is considered to be 0.8 ~\cite{PhysRevB.83.104404}. Using these parameters, $g_d \approx 2.54 \times 10^{-4}$ and $\epsilon_{co} \approx 2.53 \times 10^{-3}$ is estimated. Hence, in the region of $\epsilon \ll \epsilon_{co}$, the dipolar interaction can become relevant to alter the exponents to the mean-field value~\cite{PhysRevB.51.12585}. However, since the investigated region of $\epsilon$ for critical analysis is above $\epsilon_{co}$, it negates the possibility of dipolar interaction effects in the estimated critical exponents. Considering these arguments, the deviation from 3D Heisenberg (or any other theoretical model) indicates extended-type exchange interactions or magnetic inhomogeneity, i.e., the presence of complex exchange interactions in the system~\cite{KAUL19855}.

Since the estimated critical exponents do not follow any single universal model, it is required to check whether they eventually fall into any universal class while approaching the asymptotic region. Accordingly, reduced temperature ($\epsilon$) dependent effective critical exponents ($\beta_{eff}$ and $\gamma_{eff}$) are calculated using the equation~\cite{PhysRevB.79.214426}:
\begin{equation}
\beta_{eff}(\epsilon)=\frac{d\left[ lnM_S(\epsilon)\right ]}{d(ln\epsilon)}, \; \gamma_{eff}(\epsilon)=\frac{d \left[ ln \chi_0^{-1}(\epsilon)\right ]}{d(ln\epsilon)}
\end{equation}
Fig.~\ref{Fig:Griffiths Phase}(c) and (d) indicate that both exponents show non-monotonic behaviour with $\epsilon$. $\beta_{eff}$ has a dip at $\epsilon$= -0.007 and a peak at $\epsilon$= -0.004, while $\gamma_{eff}$ exhibits a peak at $\epsilon$= 0.006 before approaching the asymptotic limit ($\epsilon \to 0$). The $\beta_{eff}$ and $\gamma_{eff}$, at the lowest measured $\epsilon_{min}$, do not match any theoretically predicted universality class and fail to generate two different branches through the scaling equation plot. This may occur for three reasons: (i) $\epsilon_{min}$ does not fall in the asymptotic region, and to reach the asymptotic limit, $T_{\rm C}$ needs to be approached more closely, (ii) $\epsilon_{min}$, despite being in the asymptotic regime, the effective critical exponents fail to obey the universal class when the magnetic interaction is inhomogeneous, e.g., in a disordered FM system, and (iii) the asymptotic region is below the crossover regime, where the critical exponents switch to different value~\cite{PhysRevB.79.214426, PhysRevLett.91.137202, PhysRevB.100.214419}. The last option [(iii) above] of a crossover region due to dipolar interactions has already been negated earlier, and the investigated critical regime $\lvert \epsilon \rvert \simeq 10^{-2}$ is expected to be in the asymptotic region, thus invalidating the first option [(i) above] as well. Hence, the only possibility of a non-universal critical exponent at $\epsilon_{min}$ is due to the disordered FM [option (ii) above]. Further, the $\epsilon$ dependent $\gamma_{eff}$ curve resembles that of inhomogeneous FMs~\cite{Yadav2021}. In a homogeneous FM, $\gamma_{eff}(\epsilon)$ monotonically decreases with increasing $\epsilon$, whereas for a disordered FM, it shows a peak. The observed non-monotonic behaviour, along with a peak in the $\gamma_{eff}(\epsilon)$ curve, signifies the impact of disorder on the critical exponents~\cite{PhysRevB.79.214426}. Hence, the effective critical exponent analysis further implies the presence of inhomogeneous FM due to multiple complex interactions.

To understand the range and dimensionality of the magnetic interaction, renormalization-group (RG) theory has been employed. According to RG theory, the exchange interaction decays spatially as $J(r)\propto 1/r^{d+\sigma}$, where $r$, $d$, and $\sigma$ are distance, dimension of spin interaction, and measure of the range of exchange interaction, respectively~\cite{PhysRevB.108.094429}. In 3D magnetic system, when $\sigma>2$, $J(r)$ decays faster than $r^{-5}$ and interactions belongs to 3D Heisenberg model. The mean-field theory follows the interaction; $J(r)$ decreases slower than $r^{-4.5}$, with $\sigma<1.5$. For $1.5<\sigma<2$, the critical exponents belong to a different universality class based on the value of $\sigma$~\cite{PhysRevB.110.064407}. The $\sigma$ can be estimated following the equation:
\begin{equation}{\resizebox{\hsize}{!}{$\gamma=1 + \frac{4}{d}\left( \frac{n+2}{n+8}\right)\Delta\sigma + \frac{8(n+2)(n-4)}{d^2(n+8)^2} \times \left[ 1 + \frac{2G\left(\frac{d}{2}\right)(7n+20)}{(n-4)(n+8)} \right]\Delta \sigma^2$}}
\label{eq.RG}
\end{equation}
where $\Delta\sigma=(\sigma-\frac{d}{2}),  G(\frac{d}{2})=3-\frac{1}{4}(\frac{d}{2})^2$. Using Eq.~\ref{eq.RG}, along with other scaling relations $\alpha=2-\nu d$, and $2\beta=2-\alpha-\gamma$, the $d$, $n$, and $\sigma$, which yield the critical exponent values close to the experimentally obtained one, are determined~\cite{PhysRevLett.29.917, PhysRevB.107.094104}. For several sets of \{$d:n$\}, $\sigma$ has been adjusted [Table~S2 of SM~\cite{SM}] to obtain the closest value of critical exponents. For $d:n= 3:2$ and $\sigma$= 1.95, RG yielded $\gamma$= 1.352 and $\beta$= 0.36, which matches well with the experimentally determined values. The value of $\sigma$ thus obtained results in $J(r)\propto r^{-4.95}$, which suggests the localised nature of interactions of Mn spins. The spatial dependence of $J(r)$ is very close to that predicted in the 3D Heisenberg model, although a small deviation remains~\cite{PhysRevB.105.024419}. This deviation in $J(r)$ from the 3D Heisenberg model, along with the estimated value of $d:n$, indicating a 3D XY model, can be attributed to a different universality class owing to the presence of more than one magnetic interaction, i.e. magnetic inhomogeneity.  Further, the correlation length $\xi(T) \propto |T_{\rm C}-T|^{-\nu}$ corresponding to exchange interaction is associated with $\sigma$ via correlation length exponent, $\nu$, through the relation $\nu=\frac{\gamma}{\sigma}$~\cite{PhysRevB.65.064443}. Therefore, the correlation length exponent ($\nu$) and related reduced exponents $\left( \frac{\beta}{\nu}, \frac{\gamma}{\nu} \right)$ have been calculated and compared with well-established theoretical models [Table~S2 of SM~\cite{SM}]. The unique reduced critical exponents differ from all the known conventional theoretical models, suggesting an average and effective correlation length owing to the presence of multiple correlation lengths, corresponding to multiple $\nu$. Thus, the critical analysis reveals that, along with the primary exchange interaction responsible for $T_{\rm C} \sim$313~K, secondary exchange interactions are present in the system. These findings are corroborated by the d$M$/d$T$ curve [Fig.~\ref{Fig:Critical}(b)], where, along with a primary dip at $T_{\rm C}$, there exists another anomaly near $T_{\rm P} \sim$344~K.

The possibility of additional secondary magnetic interactions, indicated by the d$M(T)$/d$T$ curve at low applied field ($H=$25~Oe) and the critical analysis, has been investigated through the inverse susceptibility curve, $\chi^{-1}(T)$ [Fig.~\ref{Fig:Griffiths Phase}(e)]. In the compound, when the primary magnetic interactions are destabilized above $T_{\rm C}$, there remain ferromagnetic clusters because of the secondary interactions. The $\chi^{-1}(T)$, which is expected to follow the linear Curie-Weiss (CW) law above $T_{\rm C}$, due to these clusters, the compound can not become pure paramagnetic even above $T_{\rm C}$, exhibiting a nonlinear $\chi^{-1}(T)$ behaviour with a prominent downturn in the curve. These phenomena are called the Griffiths singularity~\cite{PhysRevLett.96.167201, x9wx-vcwv}. The high-temperature $\chi^{-1}(T)$ is fitted with the CW law $\chi^{-1} = 8(T-\theta_{\rm CW})/\mu_{eff}^2$, with CW temperature $\theta_{\rm CW}$= 344~K and effective moment $\mu_{eff}$= 4.04 $\mu_B$/f.u. [Fig.~S6(a) of SM~\cite{SM}]. The downturn of the curve from CW behaviour started below the characteristic temperature $T_{\rm G} \sim$358~K, which is the onset of FM correlation in the PM matrix, far above  $T_{\rm C}$, akin to the Griffiths phase (GP) phenomena~\cite{CHAKRABORTY2024173215}. This downturn in the $\chi^{-1}(T)$ curve may also appear due to the magnetic precursor effect, where ferromagnetic domains start forming far above $T_{\rm C}$, which are non-Griffiths-like clusters~\cite{PhysRevB.101.144433}. To distinguish the Griffiths phase from these clusters, field-dependent $\chi^{-1}(T)$ has been observed. Fig.~\ref{Fig:Griffiths Phase}(e) reveals that, upon increasing the applied magnetic field, this downturn gradually diminishes and finally disappears for the applied field of $H \geq$300~Oe [Fig.~S6(b) of SM~\cite{SM}]. This phenomenon is the hallmark of the Griffiths phase. Thus, it firmly establishes the Griffiths phase behaviour in Al$_2$MnCu for 315 $\leq T \leq$ 358~K, where the downward curvature of ${\chi}^{-1}$(T) disaapears because the contribution from the paramagnetic matrix overwhelms that of ferromagnetic clusters at high magnetic field~\cite{PhysRevB.109.224405}.

Further, the Griffiths phase is characterized through the power law, $\chi^{-1}(T) \propto (T-T_{\rm C}^{\rm R})^{1-\lambda}$ (0$\leq \lambda<$1), where $T_{\rm C}^{\rm R}$ is the critical temperature of the FM clusters and  $\lambda$ is the magnetic susceptibility exponent which becomes zero in PM region, and in the GP region it appears to be $0<\lambda_{GP}<1$~\cite{PhysRevB.77.064424}. The fitting of $\chi^{-1}$ vs $(T/T_{\rm C}^{\rm R}-1)$ curve, represented in log$_{10}$-log$_{10}$ scale for the temperature region $T \leq 346$~K and $T \geq 365$~K yielded  $\lambda_{GP}\sim$0.93(1) and $\lambda_{PM}\sim$0.006(2) respectively, with $T_{\rm C}^{\rm R}$=344~K [Fig.~\ref{Fig:Griffiths Phase}(g)]. Thus, the non-zero value of $\lambda_{GP}$ along with $T_{\rm C}^{\rm R}$, which becomes close to $\theta_{\rm CW}$, yet larger than $T_{\rm C}$, confirms Griffths phase in Al$_2$MnCu. This secondary phase has formed due to the presence of stronger secondary exchange interactions and persists well above the compound's bulk $T_{\rm C}$. Due to this higher interaction strength, the $\theta_{\rm CW}$ value has shifted to a higher value than $T_{\rm C}$. Further, this value of $\theta_{\rm CW}\sim$344~K also matches well with the ordering temperature $T_{\rm C}^{\rm R}$ of these clusters, and makes an imprint in the d$M$/d$T$ curve with a kink at the same temperature of $\sim$344~K, corroborating with the $M(T)$ measurement [Fig.~\ref{Fig:Critical}(a) and (b)] at lower applied field. All these observations firmly establish that the Griffiths phase of small ferromagnetic clusters exists with a higher ordering temperature, hence persists beyond the bulk $T_{\rm C}$.

Since the Griffiths phase exists in a low-field region and gets suppressed under application of a higher field, thermoremmanent magnetization ($M_{TRM}(T)$) measurement has been carried out with zero applied field in a specified protocol, where the compound is cooled down under an applied field, and the magnetization data is measured in a heating cycle with that field set to zero~\cite{PhysRevB.105.214407}. Fig.~\ref{Fig:Griffiths Phase}(f) shows the $M_{TRM}(T)$ curve with the y-axis on a log$_{10}$-scale plot, where an anomaly is observed between 344$<T<$358~K, making a secondary dip in the d$M_{TRM}$/d$T$ curve [Fig.~\ref{Fig:Griffiths Phase}(f)~(inset)], further substantiates the presence of the Griffiths phase in this compound.

\subsubsection{ac susceptibility}
\begin{figure}
  \centering
  \includegraphics[width=\columnwidth]{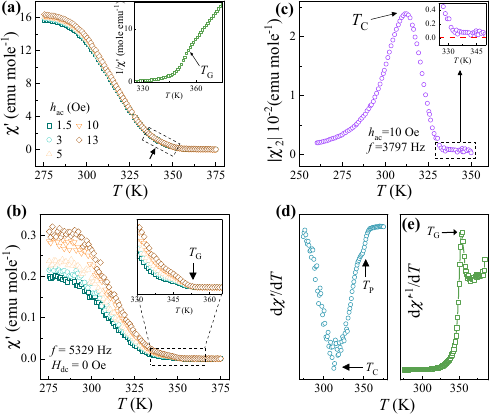}
  \caption{(a) Driving field ($h_{\rm ac}$) dependent real part ($\chi'(T)$) of ac susceptibility. Inset shows the $\chi'^{-1}(T)$ vs. $T$ plot for $h_{\rm ac}$= 10~Oe. (b) $h_{\rm ac}$ dependent imaginary part ($\chi''(T)$) of ac susceptibility. The inset is a zoomed-in figure in the temperature region around $T_{\rm G}$. (c) Temperature variation of the real part of second-harmonic ac susceptibility. The inset is the same data above $T_{\rm C}$. The first-order derivative of $\chi'$ and $\chi'^{-1}$; i.e, d$\chi'$/d$T$ (d) and  d$\chi'^{-1}$/d$T$ (e) respectively, with respect to temperature.
  }
  \label{Fig:AC}
  \end{figure}

A number of different dc magnetic measurements thus establish that the Griffiths phase in Al$_2$MnCu could be detected through low-field measurements. Since the driving field of ac magnetic measurements is generally quite small, different types of ac susceptibility ($\chi_{\rm ac}(T)$) measurements with varying frequencies ($f$) and driving fields ($h_{ac}$), as well as employing different external dc magnetic fields ($H_{dc}$) have been carried out. A careful observation of ac susceptibility data measured at all the $h_{ac}$, with $f=5329~$Hz at zero applied dc field, indeed reveals a small anomaly in the temperature range $T_{\rm C}<T<T_{\rm G}$, both in the $\chi'(T)$ and $\chi''(T)$ curves [Fig.~\ref{Fig:AC}(a) and (b)]. $\chi''(T)$ exhibits a non-zero value below $T_{\rm G}$ and follows different path upon variation of $h_{ac}$ [Fig.~\ref{Fig:AC}(b)]. Since $\chi''(T)$ is always expected to be zero in the truly paramagnetic region, here, the finite non-zero value of $\chi''(T)$ indicates the absorption of energy due to the formation of ferromagnetic clusters, far above $T_{\rm C}$~\cite{Topping_2019}. The $h_{ac}$-dependent splitting of the $\chi''(T)$ curve below $T_{\rm G}$ appears due to the different response of ferromagnetic clusters. Here, above $T_{\rm G}$, where non-interacting spins follow the driving ac field, no energy loss appears, $\chi''(T)$ essentially becomes zero, indicating a true paramagnetic phase. Further, the inset of Fig.~\ref {Fig:AC}(a) presents $\chi'^{-1}(T)$, revealing a temperature-dependent downturn, similar to that observed in dc $\chi^{-1}(T)$ [Fig.~\ref{Fig:Griffiths Phase}(e)]. More field-dependent $\chi'^{-1}(T)$ behaviour is discussed in Sec.~S3 of SM~\cite{SM}. The derivative of both $\chi'(T)$ and $\chi'^{-1}(T)$ provides a more definitive manifestation of change in slope in ac susceptibility data [Fig.~\ref{Fig:Griffiths Phase}(d) and (e)]. Similar features are observed in $f$-dependent ac susceptibility curves also [Fig.~S8(b) and (d) of SM~\cite{SM}].

The weak anomaly for the secondary phase is further investigated by the nonlinear ac measurements, which reveal the magnetic correlation of the system. In a purely paramagnetic state, where no spin correlation exists, the second harmonic of ac susceptibility $\chi_{2}$ vanishes, whereas it emerges owing to the symmetry-breaking field of FM spin clusters~\cite{Chakraborty_2024}. The $\chi_{2}(T)$ measurement on Al$_2$MnCu was carried out in the temperature range 260$-$350~K. In Fig~\ref{Fig:AC}(b), the peak of $\chi_{2}(T)$ curve is related to the FM to PM phase transition ($T_{\rm C}$) ~\cite{Nigam_2011}; the small but finite non-zero value above $T_{\rm C}$ implies the formation of FM clusters in the PM region~\cite{PhysRevB.81.024431}. To find out whether $\chi_{2}(T)$ is associated with spontaneous magnetization ($M_{\rm SP}$), isothermal magnetizations at $T>T_{\rm C}$ are investigated through the Arrott plot ($M^2$vs.$H/M$). The positive intercept on the $M^2$ axis provides $M_{\rm SP}$, whereas Fig.~\ref{Fig:Griffiths Phase}(h) shows no positive intercept on $M^2$, indicating no $M_{\rm SP}$ present above $T_{\rm C}$ in this compound. Thus, $\chi_{2}(T)$ and the Arrott plot support the emergence of the Griffiths phase above $T_{\rm C}$~\cite{GIRI2014181}.


Thus, all the experimentally obtained dc- and ac-magnetic data presented here unambiguously establish the formation of the Griffiths phase due to the enhancement of local magnetic ordering temperature over the bulk Curie temperature. Since Griffiths phase formation is associated with the presence of non-interacting magnetic clusters with short-range magnetic interaction, the additional interaction is likely caused by atomic disorder, which is sparsely present in the material. In Al$_2$MnCu, as the Al-Cu disorder is only about 5\%, it is worthy to investigate the role of this disorder in the origin of the enhanced Curie temperature for such isolated clusters.



\section{Theoretical Results}

To understand the magnetic behavior of the ordered and disordered phases, we used the procedure described in the methods section. In our $ab$ $initio$ calculations, we find that the magnetic moments on Mn atoms in different phases vary. For pristine L2$_1$, B2 and B2+12.5 \% disordered phases, the moments were $\sim$ 2.39, 2.15 and 2.25 $\mu_B$ respectively. Our Wannier-based TB successfully captures this trend, with Mn moments to be 2.55, 2.24 and 1.95 $\mu_B$ for pristine L2$_1$, B2 and B2+12.5 \% disordered phases, respectively. Similarly, the charge obtained in $ab$ $initio$ calculations on Mn $d$ is $\sim$ 5.2 $e$ is also reproduced by our TB model. This analysis, along with reasonably good agreement between our TB model with spin-polarized band structure obtained from $ab$ $initio$ (not shown here), gave us the confidence to proceed further with the analysis of magnetic interaction in the next step.

\begin{figure}[h]
  \centering
  \includegraphics[width=\columnwidth]{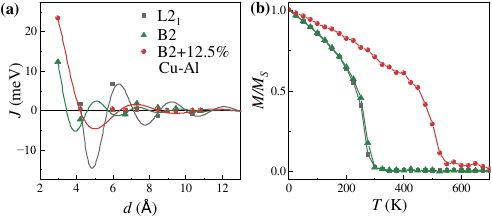}
  \caption{(a) Exchange interactions for L2$_1$ order, B2-type disorder and B2+12.5\% Cu-Al disorder structures as a function of Mn-Mn distance. The lines are fitted with a function of RKKY form and are a guide to the eye, and (b) Plot for temperature dependence of normalised magnetisation curve obtained using Monte-Carlo simulations.}
  \label{Fig:Theory}
  \end{figure}

The effective Mn-Mn magnetic interactions obtained are shown in the plot in Fig.~\ref{Fig:Theory}(a). From this figure, one can clearly observe the oscillatory nature of the exchange interaction for L2$_1$ ordered and B2 disordered phases. The solid line in this figure is a fit of these interactions to an RKKY-type function with variation in  Mn-Mn distance. It indicates that the magnetic exchange interaction is in RKKY form.

In the pristine L2$_1$ structure, there are no Mn nearest neighbors at 2.99~\AA{}. Instead, the magnetic ordering is dominated by the second-neighbor interaction, with six Mn neighbors at 5.98~\AA{} and a strong ferromagnetic (FM) exchange interaction of $\sim$6.76~meV. Although further coordination shells contain a larger number of Mn neighbors-for example, 24 neighbors at 7.32~\AA{} and another 24 at 9.45~\AA{}-their corresponding exchange interactions are considerably weaker and alternate between FM and antiferromagnetic (AFM) character. Thus, the ferromagnetic ground state in the pristine L2$_1$ phase is stabilized primarily by intermediate-range exchange interactions rather than by direct nearest-neighbor Mn--Mn coupling.

The introduction of B2-type disorder substantially modifies the Mn--Mn exchange network. Four Mn nearest neighbors are introduced at 2.99~\AA{}, with a strong FM exchange interaction of $\sim$12.34~meV. At the same time, the coordination of the second-neighbor shell at 5.98~\AA{} is reduced, while additional Mn coordination shells emerge at 6.68~\AA{} (16 neighbors) and 8.96~\AA{} (20 neighbors). Despite the nearly twofold enhancement of the nearest-neighbor exchange relative to the dominant interaction in pristine L2$_1$, the second-neighbor interaction at 5.98~\AA{} is strongly suppressed from 6.76~meV to a negligibly small value. Moreover, some of the newly introduced coordination shells exhibit AFM interactions, including $\sim$ -2.3~meV at 4.23~\AA{} and $\sim$ -1.1~meV at 6.68~\AA{}. Consequently, the enhanced nearest-neighbor FM interaction is partially compensated by the suppression of the intermediate-range FM exchange network and the emergence of competing AFM interactions.

For the B2+12.5\% Cu--Al disorder structure, the local Mn coordination remains similar to that of the B2 phase, with four nearest-neighbor Mn atoms at approximately 2.99~\AA{}. However, the corresponding nearest-neighbor exchange interaction increases substantially to $\sim$23.50~meV, nearly twice that of the B2 phase and more than three times that of the dominant exchange interaction in pristine L2$_1$. More importantly, the exchange interactions beyond the nearest-neighbor shell become negligibly small, falling predominantly within the sub-meV range. This indicates that the longer-range exchange pathways are strongly suppressed by the Cu--Al anti-site disorder, resulting in a magnetic interaction network that is dominated by the nearest-neighbor Mn--Mn coupling. This result corroborates the magnetic critical analysis [Sec.~\ref{dc magnetization}], which reveals a faster spatial decay of the exchange interaction, supporting a localized nearest-neighbor interaction.

Using Monte Carlo simulations based on the Heisenberg Hamiltonian described in the Methods section, we calculated the FM-to-paramagnetic phase transition temperatures for all three structural configurations. The exchange parameters obtained above were used as inputs to the simulations, and the resulting temperature-dependent magnetization curves are shown in Fig.~\ref{Fig:Theory}(b). Interestingly, despite the substantial changes in the nature of underlying exchange-interactions, the calculated Curie temperature of the B2 phase is nearly identical to that of the pristine L2$_1$ phase. In contrast, the introduction of 12.5\% Cu--Al anti-site disorder into the B2 structure leads to a pronounced enhancement of the Curie temperature, which is nearly twice that of the L2$_1$ and B2 phases. This enhancement can be attributed primarily to the substantial strengthening of the nearest-neighbor Mn--Mn exchange interaction and the resulting dominance of short-range ferromagnetic coupling. Analysis of Mn-$d$ orbitals contributions to magnetic interactions reveal that the major contribution to interactions comes from non-axial $d_{xy}$, $d_{yz}$ and $d_{xz}$ orbitals. Compared to L2$_1$ and B2 phases, these contribution are enhanced by an order of magnitude in B2+12.5\% phases leading to a higher Curie temperature for this case. This may a direct consequence of relatively stronger hybridization of nearest neighbor Mn $d$ electrons with conduction cloud of nearly placed non-magnetic atoms.

\section{Discussion and Conclusion}
Here, the compound inherently forms with $\sim$50\% intra-symmetry-site B2-type disorder associated with the magnetic Mn atom, along with a minor $\sim$5\%  inter-symmetry-site disorder of non-magnetic Cu and Al atoms. Magnetic critical analysis indicates the presence of complex magnetic interactions in the compound. Since the magnetic exchange interaction is responsible for the magnetic ordering temperature, due to the presence of complex secondary interactions, another secondary phase appears with a higher ordering temperature, reflected as a Griffiths phase.

Typically, structural disorder suppresses the bulk Curie temperature ($T_{\rm C}$), while localized ordered clusters retain a higher $T_{\rm C}$ to form a Griffiths phase~\cite{PhysRevB.81.024431}. Conversely, our theoretical calculation reveals that the small inter-symmetry-site disorder involving the non-magnetic atom Cu-Al elevates the ordering temperature, which appears as a Griffiths phase in Al$_2$MnCu, whereas the higher degree of intra-symmetry-site B2-type disorder, even associated with the magnetic atom Mn, does not affect the $T_{\rm C}$. This aligns with the previous report on another anti-Heusler Al$_2$MnFe, where magnetism remains unaffected by the B2-type disorder as the Mn atoms are confined to the same octahedral-symmetry site in the 4$a$ and 4$b$ Wyckoff positions~\cite{463q-ph2y}. Furthermore, in the full-Heusler Cu$_2$MnAl, the magnetic atom Mn at the 4$b$ site is surrounded by Cu atoms as the nearest-neighbor element at 8$c$ sites, and the compound exhibits a high $T_{\rm C}$ of 600~K~\cite{AniruddhaDeb_2000}. A similar atomic arrangement is observed here, when a small fraction of the Cu atoms occupy the 8$c$ site as a nearest neighbor of the Mn (4$b$) atom at a distance of 2.57 \AA \, [Fig.~\ref{Fig:structure}(h)] due to the inter-symmetry-site Cu-Al disorder, and the theoretical calculation yields a higher $T_{\rm C}$. Although it does not take into account the exact 5\% disorder, the calculation with 12.5\% disorder supports the experimentally found local enhancement of ordering temperature in the anti-Heusler alloy Al$_2$MnCu.

In conclusion, this work reports a counterintuitive role of anti-site disorder in magnetic properties. It reveals that, instead of depressing, the inter-symmetry-site disorder involving non-magnetic atoms enhances the local magnetic ordering temperature, whereas intra-symmetry-site disorder even associated with magnetic atoms does not alter the $T_{\rm C}$ in the anti-Heusler alloy Al$_2$MnCu. In the pursuit of room-temperature quantum applications based on spin-dependent phenomena, these findings establish that disorder engineering can be a promising strategy to elevate magnetic ordering temperatures.

\begin{acknowledgments}
S.B. acknowledge P.D. Babu for non-linear ac susceptibility measurements and SINP, India, for the research facility. S.K.P. acknowledges the NFSG grant from BITS-Pilani, Dubai campus, which supported this research. S.D. acknowledges the National Supercomputing Mission (NSM) for providing computing resources of ‘PARAM Kamrupa’ at IIT Guwahati, which is implemented by C-DAC and supported by the Ministry of Electronics and Information Technology (MeitY) and the Department of Science and Technology (DST), Government of India.
\end{acknowledgments}

\normalem
\bibliographystyle{apsrev4-2}
\bibliography{A2MC_GP}

@article{PhysRevB.106.134423,
  title = {Antisite disorder driven cluster glass state and colossal magnetoresistance in ${\mathrm{MnSb}}_{2}{\mathrm{Se}}_{4}$},
  author = {Kumar, Rahul and Sundaresan, A.},
  journal = {Phys. Rev. B},
  volume = {106},
  issue = {13},
  pages = {134423},
  numpages = {10},
  year = {2022},
  month = {Oct},
  publisher = {American Physical Society},
  doi = {10.1103/PhysRevB.106.134423},
  url = {https://link.aps.org/doi/10.1103/PhysRevB.106.134423}
}

@article{PhysRevB.96.144423,
  title = {Magnetism out of antisite disorder in the $J=0$ compound ${\mathrm{Ba}}_{2}{\mathrm{YIrO}}_{6}$},
  author = {Chen, Q. and Svoboda, C. and Zheng, Q. and Sales, B. C. and Mandrus, D. G. and Zhou, H. D. and Zhou, J.-S. and McComb, D. and Randeria, M. and Trivedi, N. and Yan, J.-Q.},
  journal = {Phys. Rev. B},
  volume = {96},
  issue = {14},
  pages = {144423},
  numpages = {11},
  year = {2017},
  month = {Oct},
  publisher = {American Physical Society},
  doi = {10.1103/PhysRevB.96.144423},
  url = {https://link.aps.org/doi/10.1103/PhysRevB.96.144423}
}

@Article{Garmroudi2022,
author={Garmroudi, Fabian
and Parzer, Michael
and Riss, Alexander
and Ruban, Andrei V.
and Khmelevskyi, Sergii
and Reticcioli, Michele
and Knopf, Matthias
and Michor, Herwig
and Pustogow, Andrej
and Mori, Takao
and Bauer, Ernst},
title={Anderson transition in stoichiometric Fe2VAl: high thermoelectric performance from impurity bands},
journal={Nature Communications},
year={2022},
month={Jun},
day={23},
volume={13},
number={1},
pages={3599},
issn={2041-1723},
doi={10.1038/s41467-022-31159-w},
url={https://doi.org/10.1038/s41467-022-31159-w}
}

@article{doi:10.1073/pnas.1810842115,
author = {Chandra Shekhar  and Nitesh Kumar  and V. Grinenko  and Sanjay Singh  and R. Sarkar  and H. Luetkens  and Shu-Chun Wu  and Yang Zhang  and Alexander C. Komarek  and Erik Kampert  and Yurii Skourski  and Jochen Wosnitza  and Walter Schnelle  and Alix McCollam  and Uli Zeitler  and Jürgen Kübler  and Binghai Yan  and H.-H. Klauss  and S. S. P. Parkin  and C. Felser },
title = {Anomalous Hall effect in Weyl semimetal half-Heusler compounds RPtBi (R = Gd and Nd)},
journal = {Proceedings of the National Academy of Sciences},
volume = {115},
number = {37},
pages = {9140-9144},
year = {2018},
doi = {10.1073/pnas.1810842115},
URL = {https://www.pnas.org/doi/abs/10.1073/pnas.1810842115},
eprint = {https://www.pnas.org/doi/pdf/10.1073/pnas.1810842115},
}

@article{PhysRevB.108.045137,
  title = {High spin polarization in the disordered quaternary Heusler alloy FeMnVGa},
  author = {Gupta, Shuvankar and Chakraborty, Sudip and Bhasin, Vidha and Pakhira, Santanu and Dan, Shovan and Barreteau, Celine and Crivello, Jean-Claude and Jha, S. N. and Avdeev, Maxim and Greneche, Jean-Marc and Bhattacharyya, D. and Alleno, Eric and Mazumdar, Chandan},
  journal = {Phys. Rev. B},
  volume = {108},
  issue = {4},
  pages = {045137},
  numpages = {12},
  year = {2023},
  month = {Jul},
  publisher = {American Physical Society},
  doi = {10.1103/PhysRevB.108.045137},
  url = {https://link.aps.org/doi/10.1103/PhysRevB.108.045137}
}

@article{PhysRevLett.110.100401,
  title = {Realization of Spin Gapless Semiconductors: The Heusler Compound ${\mathrm{Mn}}_{2}\mathrm{CoAl}$},
  author = {Ouardi, Siham and Fecher, Gerhard H. and Felser, Claudia and K\"ubler, J\"urgen},
  journal = {Phys. Rev. Lett.},
  volume = {110},
  issue = {10},
  pages = {100401},
  numpages = {5},
  year = {2013},
  month = {Mar},
  publisher = {American Physical Society},
  doi = {10.1103/PhysRevLett.110.100401},
  url = {https://link.aps.org/doi/10.1103/PhysRevLett.110.100401}
}

@article{PhysRevLett.47.1480,
  title = {Disorder and the Optical-Absorption Edge of Hydrogenated Amorphous Silicon},
  author = {Cody, G. D. and Tiedje, T. and Abeles, B. and Brooks, B. and Goldstein, Y.},
  journal = {Phys. Rev. Lett.},
  volume = {47},
  issue = {20},
  pages = {1480--1483},
  numpages = {0},
  year = {1981},
  month = {Nov},
  publisher = {American Physical Society},
  doi = {10.1103/PhysRevLett.47.1480},
  url = {https://link.aps.org/doi/10.1103/PhysRevLett.47.1480}
}

@article{GRAF20111,
title = {Simple rules for the understanding of Heusler compounds},
journal = {Progress in Solid State Chemistry},
volume = {39},
number = {1},
pages = {1-50},
year = {2011},
issn = {0079-6786},
doi = {https://doi.org/10.1016/j.progsolidstchem.2011.02.001},
url = {https://www.sciencedirect.com/science/article/pii/S0079678611000021},
author = {Tanja Graf and Claudia Felser and Stuart S.P. Parkin}
}

@article{PhysRevB.97.184421,
  title = {Reentrant cluster glass and stability of ferromagnetism in the ${\mathrm{Ga}}_{2}\mathrm{MnCo}$ Heusler alloy},
  author = {Samanta, Tamalika and Bhobe, P. A. and Das, A. and Kumar, A. and Nigam, A. K.},
  journal = {Phys. Rev. B},
  volume = {97},
  issue = {18},
  pages = {184421},
  numpages = {7},
  year = {2018},
  month = {May},
  publisher = {American Physical Society},
  doi = {10.1103/PhysRevB.97.184421},
  url = {https://link.aps.org/doi/10.1103/PhysRevB.97.184421}
}

@Article{Samanta2022,
author={Samanta, Tamalika
and Velaga, Srihari
and Bhobe, P. A.},
title={Unraveling the Magnetic Ground State and Local Lattice Distortions in Z2XY-Type Full Heusler Compounds: An EXAFS Study},
journal={The Journal of Physical Chemistry C},
year={2022},
month={Oct},
day={20},
publisher={American Chemical Society},
volume={126},
number={41},
pages={17670-17679},
issn={1932-7447},
doi={10.1021/acs.jpcc.2c04088},
url={https://doi.org/10.1021/acs.jpcc.2c04088}
}

@article{PhysRevB.111.174417,
  title = {Al$_2\rm {MnCu}$: A magnetically ordered member of the $\rm {Heusler}$ alloy family despite having a valence electron count of 24},
  author = {Bhowmik, Soumya and Pakhira, Santanu and Choudhary, Renu and Kumar, Ravi and Urkude, Rajashri and Ghosh, Biplab and Bhattacharyya, D. and Avdeev, Maxim and Mazumdar, Chandan},
  journal = {Phys. Rev. B},
  volume = {111},
  issue = {17},
  pages = {174417},
  numpages = {11},
  year = {2025},
  month = {May},
  publisher = {American Physical Society},
  doi = {10.1103/PhysRevB.111.174417},
  url = {https://link.aps.org/doi/10.1103/PhysRevB.111.174417}
}

@article{463q-ph2y,
  title = {Disproportionate influence of site disorder on the evolution of magnetic phases in the anti-Heusler alloy ${\mathrm{Al}}_{2}\mathrm{MnFe}$},
  author = {Bhowmik, Soumya and Pakhira, Santanu and Kundu, Ashis and Reddy, V. Raghavendra and Kabir, Mukul and Mazumdar, Chandan},
  journal = {Phys. Rev. B},
  volume = {113},
  issue = {1},
  pages = {014429},
  numpages = {18},
  year = {2026},
  month = {Jan},
  publisher = {American Physical Society},
  doi = {10.1103/463q-ph2y},
  url = {https://link.aps.org/doi/10.1103/463q-ph2y}
}

@article{Avdeev:ks5608,
author = "Avdeev, Maxim and Hester, James R.",
title = "{ECHIDNA: a decade of high-resolution neutron powder diffraction at OPAL}",
journal = "Journal of Applied Crystallography",
year = "2018",
volume = "51",
number = "6",
pages = "1597--1604",
month = "Dec",
doi = {10.1107/S1600576718014048},
url = {https://doi.org/10.1107/S1600576718014048},
}

@article{RODRIGUEZCARVAJAL199355,
title = {Recent advances in magnetic structure determination by neutron powder diffraction},
journal = {Physica B: Condensed Matter},
volume = {192},
number = {1},
pages = {55-69},
year = {1993},
issn = {0921-4526},
doi = {https://doi.org/10.1016/0921-4526(93)90108-I},
url = {https://www.sciencedirect.com/science/article/pii/092145269390108I},
author = {Juan Rodríguez-Carvajal},
}

@article{ArrottPlot,
  title = {Criterion for Ferromagnetism from Observations of Magnetic Isotherms},
  author = {Anthony Arrott},
  journal = {Phys. Rev. },
  volume = {108},
  issue = {6},
  pages = {1394},
  numpages = {1},
  year = {1957},
  month = {December},
  publisher = {American Physical Society},
  doi = {10.1103/PhysRev.108.1394},
  url = {https://doi.org/10.1103/PhysRev.108.1394}
}

@article{BANERJEE196416,
title = {On a generalised approach to first and second order magnetic transitions},
author = {Banerjee, Bhabani K.},
journal = {Physics Letters},
volume = {12},
number = {1},
pages = {16-17},
year = {1964},
issn = {0031-9163},
doi = {https://doi.org/10.1016/0031-9163(64)91158-8},
url = {https://www.sciencedirect.com/science/article/pii/0031916364911588},

}

@article{PhysRevB.105.214407,
  title = {Origin of the Griffiths phase and correlation with the magnetic phase transition in the nanocrystalline manganite ${\mathrm{La}}_{0.4}{({\mathrm{Ca}}_{0.5}{\mathrm{Sr}}_{0.5})}_{0.6}{\mathrm{MnO}}_{3}$},
  author = {Saha, Suvayan and Dutta, Apurba and Gupta, Shuvankar and Bandyopadhyay, Sudipta and Das, I.},
  journal = {Phys. Rev. B},
  volume = {105},
  issue = {21},
  pages = {214407},
  numpages = {16},
  year = {2022},
  month = {Jun},
  publisher = {American Physical Society},
  doi = {10.1103/PhysRevB.105.214407},
  url = {https://link.aps.org/doi/10.1103/PhysRevB.105.214407}
}

@article{PhysRevLett.19.786,
  title = {Approximate Equation of State For Nickel Near its Critical Temperature},
  author = {Arrott, Anthony and Noakes, John E.},
  journal = {Phys. Rev. Lett.},
  volume = {19},
  issue = {14},
  pages = {786--789},
  numpages = {0},
  year = {1967},
  month = {Oct},
  publisher = {American Physical Society},
  doi = {10.1103/PhysRevLett.19.786},
  url = {https://link.aps.org/doi/10.1103/PhysRevLett.19.786}
}

@book{stanley1971introduction,
  title     = {Introduction to Phase Transitions and Critical Phenomena},
  author    = {Stanley, H. Eugene},
  year      = {1971},
  publisher = {Oxford University Press},
  address   = {New York},
  series    = {International Series of Monographs on Physics},
  isbn      = {9780195053166}
}

@article{KFPlot,
  title     = {Detailed Magnetic Behavior of Nickel Near its Curie Point},
  author    = {Kouvel, James S. and Fisher, Michael E.},
  journal   = {Physical Review},
  volume    = {136},
  issue     = {5A},
  pages     = {A1626--A1632},
  year      = {1964},
  month     = {Dec},
  publisher = {American Physical Society},
  doi       = {10.1103/PhysRev.136.A1626},
  url       = {https://link.aps.org/doi/10.1103/PhysRev.136.A1626}
}

@article{10.1063/1.1696618,
    author = {Widom, B.},
    title = {Equation of State in the Neighborhood of the Critical Point},
    journal = {The Journal of Chemical Physics},
    volume = {43},
    number = {11},
    pages = {3898-3905},
    year = {1965},
    month = {12},
    issn = {0021-9606},
    doi = {10.1063/1.1696618},
    url = {https://doi.org/10.1063/1.1696618},
}

@article{PhysRevB.109.094408,
  title = {Critical behavior, magnetic phase diagram, and magnetic entropy change of ${\mathrm{MnSb}}_{2}{\mathrm{Te}}_{4}$},
  author = {Yang, Xiaojun and Pan, Junxiao and He, Xiangyu and Chu, Dongliang},
  journal = {Phys. Rev. B},
  volume = {109},
  issue = {9},
  pages = {094408},
  numpages = {11},
  year = {2024},
  month = {Mar},
  publisher = {American Physical Society},
  doi = {10.1103/PhysRevB.109.094408},
  url = {https://link.aps.org/doi/10.1103/PhysRevB.109.094408}
}

@article{PhysRevB.32.7209,
  title = {Nonasymptotic critical behavior from field theory at $d=3$: The disordered-phase case},
  author = {Bagnuls, C. and Bervillier, C.},
  journal = {Phys. Rev. B},
  volume = {32},
  issue = {11},
  pages = {7209--7231},
  numpages = {0},
  year = {1985},
  month = {Dec},
  publisher = {American Physical Society},
  doi = {10.1103/PhysRevB.32.7209},
  url = {https://link.aps.org/doi/10.1103/PhysRevB.32.7209}
}

@article{KAUL19855,
title = {Static critical phenomena in ferromagnets with quenched disorder},
journal = {Journal of Magnetism and Magnetic Materials},
volume = {53},
number = {1},
pages = {5-53},
year = {1985},
issn = {0304-8853},
doi = {https://doi.org/10.1016/0304-8853(85)90128-3},
url = {https://www.sciencedirect.com/science/article/pii/0304885385901283},
author = {S.N. Kaul},
}

@article{PhysRevB.65.064443,
  title = {Critical magnetic properties of disordered polycrystalline ${\mathrm{Cr}}_{75}{\mathrm{Fe}}_{25}$ and ${\mathrm{Cr}}_{70}{\mathrm{Fe}}_{30}$ alloys},
  author = {Fischer, S. F. and Kaul, S. N. and Kronm\"uller, H.},
  journal = {Phys. Rev. B},
  volume = {65},
  issue = {6},
  pages = {064443},
  numpages = {12},
  year = {2002},
  month = {Jan},
  publisher = {American Physical Society},
  doi = {10.1103/PhysRevB.65.064443},
  url = {https://link.aps.org/doi/10.1103/PhysRevB.65.064443}
}

@article{PhysRevB.65.144520,
  title = {Critical exponents and equation of state of the three-dimensional Heisenberg universality class},
  author = {Campostrini, Massimo and Hasenbusch, Martin and Pelissetto, Andrea and Rossi, Paolo and Vicari, Ettore},
  journal = {Phys. Rev. B},
  volume = {65},
  issue = {14},
  pages = {144520},
  numpages = {21},
  year = {2002},
  month = {Apr},
  publisher = {American Physical Society},
  doi = {10.1103/PhysRevB.65.144520},
  url = {https://link.aps.org/doi/10.1103/PhysRevB.65.144520}
}

@article{PhysRevB.21.3976,
  title = {Critical exponents from field theory},
  author = {Le Guillou, J. C. and Zinn-Justin, J.},
  journal = {Phys. Rev. B},
  volume = {21},
  issue = {9},
  pages = {3976--3998},
  numpages = {0},
  year = {1980},
  month = {May},
  publisher = {American Physical Society},
  doi = {10.1103/PhysRevB.21.3976},
  url = {https://link.aps.org/doi/10.1103/PhysRevB.21.3976}
}

@article{PhysRevB.99.214414,
  title = {Complex exchange mechanism driven ferromagnetism in half-metallic Heusler ${\mathrm{Co}}_{2}\mathrm{TiGe}$: Evidence from critical behavior},
  author = {Roy, Shubhankar and Khan, Nazir and Singha, Ratnadwip and Pariari, Arnab and Mandal, Prabhat},
  journal = {Phys. Rev. B},
  volume = {99},
  issue = {21},
  pages = {214414},
  numpages = {8},
  year = {2019},
  month = {Jun},
  publisher = {American Physical Society},
  doi = {10.1103/PhysRevB.99.214414},
  url = {https://link.aps.org/doi/10.1103/PhysRevB.99.214414}
}

@article{PhysRevLett.30.559,
  title = {Dipolar Interactions at Ferromagnetic Critical Points},
  author = {Fisher, Michael E. and Aharony, Amnon},
  journal = {Phys. Rev. Lett.},
  volume = {30},
  issue = {12},
  pages = {559--562},
  numpages = {0},
  year = {1973},
  month = {Mar},
  publisher = {American Physical Society},
  doi = {10.1103/PhysRevLett.30.559},
  url = {https://link.aps.org/doi/10.1103/PhysRevLett.30.559}
}

@article{PhysRevB.51.12585,
  title = {Asymptotic critical behavior of Ni},
  author = {Seeger, M. and Kaul, S. N. and Kronm\"uller, H. and Reisser, R.},
  journal = {Phys. Rev. B},
  volume = {51},
  issue = {18},
  pages = {12585--12594},
  numpages = {0},
  year = {1995},
  month = {May},
  publisher = {American Physical Society},
  doi = {10.1103/PhysRevB.51.12585},
  url = {https://link.aps.org/doi/10.1103/PhysRevB.51.12585}
}

@article{PhysRevB.8.3323,
  title = {Critical Behavior of Magnets with Dipolar Interactions. I. Renormalization Group near Four Dimensions},
  author = {Aharony, Amnon and Fisher, Michael E.},
  journal = {Phys. Rev. B},
  volume = {8},
  issue = {7},
  pages = {3323--3341},
  numpages = {0},
  year = {1973},
  month = {Oct},
  publisher = {American Physical Society},
  doi = {10.1103/PhysRevB.8.3323},
  url = {https://link.aps.org/doi/10.1103/PhysRevB.8.3323}
}

@article{RIED1992259,
title = {Experimental hints for crossover behaviour in the ferromagnetic phase transition of Fe14Nd2B},
journal = {Journal of Magnetism and Magnetic Materials},
volume = {116},
number = {1},
pages = {259-266},
year = {1992},
issn = {0304-8853},
doi = {https://doi.org/10.1016/0304-8853(92)90170-S},
url = {https://www.sciencedirect.com/science/article/pii/030488539290170S},
author = {K. Ried and D. Köhler and H. Kronmüller},
}

@article{PhysRevB.79.214426,
  title = {Critical behavior at paramagnetic to ferromagnetic phase transition in ${\text{Pr}}_{0.5}{\text{Sr}}_{0.5}{\text{MnO}}_{3}$: A bulk magnetization study},
  author = {Pramanik, A. K. and Banerjee, A.},
  journal = {Phys. Rev. B},
  volume = {79},
  issue = {21},
  pages = {214426},
  numpages = {7},
  year = {2009},
  month = {Jun},
  publisher = {American Physical Society},
  doi = {10.1103/PhysRevB.79.214426},
  url = {https://link.aps.org/doi/10.1103/PhysRevB.79.214426}
}

@article{PhysRevLett.91.137202,
  title = {Quenched Disorder and the Critical Behavior of a Partially Frustrated System},
  author = {Perumal, A. and Srinivas, V. and Rao, V. V. and Dunlap, R. A.},
  journal = {Phys. Rev. Lett.},
  volume = {91},
  issue = {13},
  pages = {137202},
  numpages = {4},
  year = {2003},
  month = {Sep},
  publisher = {American Physical Society},
  doi = {10.1103/PhysRevLett.91.137202},
  url = {https://link.aps.org/doi/10.1103/PhysRevLett.91.137202}
}

@article{PhysRevB.100.214419,
  title = {Critical behavior in the half-metallic Heusler alloy ${\mathrm{Co}}_{2}\mathrm{TiSn}$},
  author = {Rahman, Azizur and Rehman, Majeed ur and Zhang, Dechen and Zhang, Min and Wang, Xiangqi and Dai, Rucheng and Wang, Zhongping and Tao, Xiaoping and Zhang, Lei and Zhang, Zengming},
  journal = {Phys. Rev. B},
  volume = {100},
  issue = {21},
  pages = {214419},
  numpages = {6},
  year = {2019},
  month = {Dec},
  publisher = {American Physical Society},
  doi = {10.1103/PhysRevB.100.214419},
  url = {https://link.aps.org/doi/10.1103/PhysRevB.100.214419}
}

@Article{Yadav2021,
author={Yadav, Kavita
and Ranaut, Dheeraj
and Mukherjee, K.},
title={Unconventional critical behaviour in weak ferromagnets Fe2-xMnxCrAl (0{\thinspace}≤{\thinspace}x{\thinspace}<{\thinspace}1)},
journal={Scientific Reports},
year={2021},
month={Sep},
day={21},
volume={11},
number={1},
pages={18742},
issn={2045-2322},
doi={10.1038/s41598-021-98377-y},
url={https://doi.org/10.1038/s41598-021-98377-y}
}

@article{PhysRevB.110.064407,
  title = {Exploring the tricritical nature and unconventional magnetoresistance in ${\mathrm{EuCd}}_{2}{\mathrm{P}}_{2}$: An in-depth study of complex material characteristics},
  author = {Rahman, Azizur and Rehman, Majeed Ur and Zhang, Yue and Zhao, Weisheng and Wang, Jianlin and Chen, Zheng and Muhammad, Zahir and Zhang, Lei},
  journal = {Phys. Rev. B},
  volume = {110},
  issue = {6},
  pages = {064407},
  numpages = {9},
  year = {2024},
  month = {Aug},
  publisher = {American Physical Society},
  doi = {10.1103/PhysRevB.110.064407},
  url = {https://link.aps.org/doi/10.1103/PhysRevB.110.064407}
}

@article{PhysRevB.109.054413,
  title = {Critical behavior in monoclinic ${\mathrm{Cr}}_{3}{\mathrm{Te}}_{4}$},
  author = {Goswami, Anirban and Ng, Nicholas and Yakubu, Emmanuel and Abeykoon, AM Milinda and Guchhait, Samaresh},
  journal = {Phys. Rev. B},
  volume = {109},
  issue = {5},
  pages = {054413},
  numpages = {10},
  year = {2024},
  month = {Feb},
  publisher = {American Physical Society},
  doi = {10.1103/PhysRevB.109.054413},
  url = {https://link.aps.org/doi/10.1103/PhysRevB.109.054413}
}

@article{PhysRevB.107.094104,
  title = {Fe-doping modulated critical behavior and magnetocaloric effect of magnetic Weyl ${\mathrm{Co}}_{3}{\mathrm{Sn}}_{2}{\mathrm{S}}_{2}$},
  author = {Hu, Jiyu and Chen, Zheng and Wang, Meiling and Wang, Min and Wang, Shouguo and Zheng, Ganhong and Liang, Xin and Lu, Wenjian and Ma, Yongqing},
  journal = {Phys. Rev. B},
  volume = {107},
  issue = {9},
  pages = {094104},
  numpages = {12},
  year = {2023},
  month = {Mar},
  publisher = {American Physical Society},
  doi = {10.1103/PhysRevB.107.094104},
  url = {https://link.aps.org/doi/10.1103/PhysRevB.107.094104}
}

@article{PhysRevLett.29.917,
  title = {Critical Exponents for Long-Range Interactions},
  author = {Fisher, Michael E. and Ma, Shang-keng and Nickel, B. G.},
  journal = {Phys. Rev. Lett.},
  volume = {29},
  issue = {14},
  pages = {917--920},
  numpages = {0},
  year = {1972},
  month = {Oct},
  publisher = {American Physical Society},
  doi = {10.1103/PhysRevLett.29.917},
  url = {https://link.aps.org/doi/10.1103/PhysRevLett.29.917}
}

@article{PhysRevB.105.024419,
  title = {Three-dimensional critical behavior and anisotropic magnetic entropy change in quasi-two-dimensional ${\mathrm{LaCrSb}}_{3}$},
  author = {Yang, Xiaojun and Pan, Junxiao and Gai, Weizhuo and Tao, Yaping and Jia, Hong and Cao, Leiming and Cao, Yan},
  journal = {Phys. Rev. B},
  volume = {105},
  issue = {2},
  pages = {024419},
  numpages = {11},
  year = {2022},
  month = {Jan},
  publisher = {American Physical Society},
  doi = {10.1103/PhysRevB.105.024419},
  url = {https://link.aps.org/doi/10.1103/PhysRevB.105.024419}
}

@article{PhysRevLett.96.167201,
  title = {Observation of a Griffiths-like Phase in the Magnetocaloric Compound ${\mathrm{Tb}}_{5}{\mathrm{Si}}_{2}{\mathrm{Ge}}_{2}$},
  author = {Magen, C. and Algarabel, P. A. and Morellon, L. and Ara\'ujo, J. P. and Ritter, C. and Ibarra, M. R. and Pereira, A. M. and Sousa, J. B.},
  journal = {Phys. Rev. Lett.},
  volume = {96},
  issue = {16},
  pages = {167201},
  numpages = {4},
  year = {2006},
  month = {Apr},
  publisher = {American Physical Society},
  doi = {10.1103/PhysRevLett.96.167201},
  url = {https://link.aps.org/doi/10.1103/PhysRevLett.96.167201}
}

@article{PhysRevB.83.104404,
  title = {Critical behavior in the inhomogeneous ferromagnet ${\mathrm{SrFe}}_{0.80}{\mathrm{Co}}_{0.20}{\mathrm{O}}_{3.0}$},
  author = {Lago, J. and Rosseinsky, M. J. and Blundell, S. J. and Battle, P. D. and Diaz, M. and Uriarte, I. and Rojo, T.},
  journal = {Phys. Rev. B},
  volume = {83},
  issue = {10},
  pages = {104404},
  numpages = {9},
  year = {2011},
  month = {Mar},
  publisher = {American Physical Society},
  doi = {10.1103/PhysRevB.83.104404},
  url = {https://link.aps.org/doi/10.1103/PhysRevB.83.104404}
}

@article{PhysRevB.109.224405,
  title = {Griffiths phase arising from local lattice distortion and spin glass above the Curie temperature in ${\mathrm{Ni}}_{2}\mathrm{MnSb}$ polycrystalline Heusler alloy},
  author = {Tian, Fanghua and Zhao, Qizhong and Guo, Jiale and Zhang, Yin and Fang, Minxia and Chang, Tieyan and Dai, Zhiyong and Zhou, Chao and Cao, Kaiyan and Yang, Sen},
  journal = {Phys. Rev. B},
  volume = {109},
  issue = {22},
  pages = {224405},
  numpages = {9},
  year = {2024},
  month = {Jun},
  publisher = {American Physical Society},
  doi = {10.1103/PhysRevB.109.224405},
  url = {https://link.aps.org/doi/10.1103/PhysRevB.109.224405}
}

@article{CHAKRABORTY2024173215,
title = {Rare coexistence of disorder-induced Griffiths phase and reentrant spin-glass state in a Heusler alloy Rh2FeAl with high Curie temperature},
journal = {Journal of Alloys and Compounds},
volume = {976},
pages = {173215},
year = {2024},
issn = {0925-8388},
doi = {https://doi.org/10.1016/j.jallcom.2023.173215},
url = {https://www.sciencedirect.com/science/article/pii/S0925838823045188},
author = {Sudip Chakraborty and Shuvankar Gupta and Chandan Mazumdar},
}

@article{x9wx-vcwv,
  title = {Evidence of Griffiths-like clustered phase and nonuniversal critical behavior in the double perovskite ${\mathrm{Er}}_{2}{\mathrm{NiMnO}}_{6}$},
  author = {Mazumdar, Dipak and Das, I.},
  journal = {Phys. Rev. B},
  volume = {112},
  issue = {21},
  pages = {214432},
  numpages = {25},
  year = {2025},
  month = {Dec},
  publisher = {American Physical Society},
  doi = {10.1103/x9wx-vcwv},
  url = {https://link.aps.org/doi/10.1103/x9wx-vcwv}
}

@article{PhysRevB.77.064424,
  title = {Griffiths phase and critical behavior in single-crystal ${\mathrm{La}}_{0.7}{\mathrm{Ba}}_{0.3}\mathrm{Mn}{\mathrm{O}}_{3}$: Phase diagram for ${\mathrm{La}}_{1\ensuremath{-}x}{\mathrm{Ba}}_{x}\mathrm{Mn}{\mathrm{O}}_{3}$ $(x\ensuremath{\leqslant}0.33)$},
  author = {Jiang, Wanjun and Zhou, XueZhi and Williams, Gwyn and Mukovskii, Y. and Glazyrin, K.},
  journal = {Phys. Rev. B},
  volume = {77},
  issue = {6},
  pages = {064424},
  numpages = {7},
  year = {2008},
  month = {Feb},
  publisher = {American Physical Society},
  doi = {10.1103/PhysRevB.77.064424},
  url = {https://link.aps.org/doi/10.1103/PhysRevB.77.064424}
}

@article{PhysRevB.101.144433,
  title = {Non-Griffiths-like cluster formation in the double-perovskite ${\mathrm{Gd}}_{2}{\mathrm{CoMnO}}_{6}$: Evidence from critical behavior},
  author = {Das, Moumita and Sarkar, Prosenjit and Mandal, Prabhat},
  journal = {Phys. Rev. B},
  volume = {101},
  issue = {14},
  pages = {144433},
  numpages = {8},
  year = {2020},
  month = {Apr},
  publisher = {American Physical Society},
  doi = {10.1103/PhysRevB.101.144433},
  url = {https://link.aps.org/doi/10.1103/PhysRevB.101.144433}
}

@article{Topping_2019,
doi = {10.1088/1361-648X/aaed96},
url = {https://doi.org/10.1088/1361-648X/aaed96},
year = {2018},
month = {nov},
publisher = {IOP Publishing},
volume = {31},
number = {1},
pages = {013001},
author = {Topping, C V and Blundell, S J},
title = {A.C. susceptibility as a probe of low-frequency magnetic dynamics},
journal = {Journal of Physics: Condensed Matter},
}

@article{Chakraborty_2024,
doi = {10.1088/1402-4896/ad36f4},
url = {https://doi.org/10.1088/1402-4896/ad36f4},
year = {2024},
month = {apr},
publisher = {IOP Publishing},
volume = {99},
number = {5},
pages = {055924},
author = {Chakraborty, Tirthankar and Sharma, Nikita and Elizabeth, Suja},
title = {Unraveling the origin of the Griffiths phase in Ho2NiMnO6: perspectives from linear and nonlinear AC susceptibility},
journal = {Physica Scripta},
}

@article{PhysRevB.81.024431,
  title = {Griffiths phase and its evolution with Mn-site disorder in the half-doped manganite ${\text{Pr}}_{0.5}{\text{Sr}}_{0.5}{\text{Mn}}_{1\ensuremath{-}y}{\text{Ga}}_{y}{\text{O}}_{3}$ ($y=0.0$, 0.025, and 0.05)},
  author = {Pramanik, A. K. and Banerjee, A.},
  journal = {Phys. Rev. B},
  volume = {81},
  issue = {2},
  pages = {024431},
  numpages = {5},
  year = {2010},
  month = {Jan},
  publisher = {American Physical Society},
  doi = {10.1103/PhysRevB.81.024431},
  url = {https://link.aps.org/doi/10.1103/PhysRevB.81.024431}
}

@article{PhysRevB.108.094429,
  title = {Field-dependent anisotropic room-temperature ferromagnetism in ${\mathrm{Cr}}_{3}{\mathrm{Te}}_{4}$},
  author = {Wang, Aina and Rahman, Azizur and Du, Zan and Zhao, Jun and Meng, Fanying and Liu, Wei and Fan, Jiyu and Ma, Chunlan and Ge, Min and Pi, Li and Zhang, Lei and Zhang, Yuheng},
  journal = {Phys. Rev. B},
  volume = {108},
  issue = {9},
  pages = {094429},
  numpages = {8},
  year = {2023},
  month = {Sep},
  publisher = {American Physical Society},
  doi = {10.1103/PhysRevB.108.094429},
  url = {https://link.aps.org/doi/10.1103/PhysRevB.108.094429}
}

@article{GIRI2014181,
title = {Emergence of Griffiths phase and glassy mixed phase in Sm0.5Ca0.5MnO3 nanomanganites},
journal = {Journal of Alloys and Compounds},
volume = {591},
pages = {181-187},
year = {2014},
issn = {0925-8388},
doi = {https://doi.org/10.1016/j.jallcom.2013.12.076},
url = {https://www.sciencedirect.com/science/article/pii/S0925838813030557},
author = {S.K. Giri and S.M. Yusuf and M.D. Mukadam and T.K. Nath},
}

@article{Nigam_2011,
doi = {10.1088/0953-8984/23/43/435702},
url = {https://doi.org/10.1088/0953-8984/23/43/435702},
year = {2011},
month = {oct},
publisher = {IOP Publishing},
volume = {23},
number = {43},
pages = {435702},
author = {Nigam, R and Pan, A V and Dou, S X},
title = {Magnetic phase diagrams based on static and dynamic magnetic behaviour in Ru-based superconducting ferromagnets},
journal = {Journal of Physics: Condensed Matter},
}

@Article{León2025,
author={Le{\'o}n, Andrea
and Autieri, Carmine
and Brumme, Thomas
and Gonz{\'a}lez, Jhon W.},
title={Hybrid d/p-wave altermagnetism in Ca3Ru2O7 and strain-controlled spin splitting},
journal={npj Quantum Materials},
year={2025},
month={Sep},
day={26},
volume={10},
number={1},
pages={98},
doi={10.1038/s41535-025-00814-y},
url={https://doi.org/10.1038/s41535-025-00814-y}
}

@article{PhysRevB.102.134410,
  title = {Effect of chemical substitution on the skyrmion phase in ${\mathrm{Cu}}_{2}{\mathrm{OSeO}}_{3}$},
  author = {Neves, Paul M. and Gilbert, Dustin A. and Ran, Sheng and Liu, I-Lin and Saha, Shanta and Collini, John and Bleuel, Markus and Paglione, Johnpierre and Borchers, Julie A. and Butch, Nicholas P.},
  journal = {Phys. Rev. B},
  volume = {102},
  issue = {13},
  pages = {134410},
  numpages = {12},
  year = {2020},
  month = {Oct},
  publisher = {American Physical Society},
  doi = {10.1103/PhysRevB.102.134410},
  url = {https://link.aps.org/doi/10.1103/PhysRevB.102.134410}
}

@article{PhysRevLett.119.157201,
  title = {Disorder-Induced Mimicry of a Spin Liquid in ${\mathrm{YbMgGaO}}_{4}$},
  author = {Zhu, Zhenyue and Maksimov, P. A. and White, Steven R. and Chernyshev, A. L.},
  journal = {Phys. Rev. Lett.},
  volume = {119},
  issue = {15},
  pages = {157201},
  numpages = {6},
  year = {2017},
  month = {Oct},
  publisher = {American Physical Society},
  doi = {10.1103/PhysRevLett.119.157201},
  url = {https://link.aps.org/doi/10.1103/PhysRevLett.119.157201}
}

@article{PhysRevB.97.165137,
  title = {Effect of anti-site disorder on magnetism in ${\mathrm{La}}_{2}{\mathrm{NiMnO}}_{6}$},
  author = {Pal, Somnath and Sharada Govinda and Goyal, Manik and Mukherjee, Soham and Pal, Banabir and Saha, Rana and Sundaresan, A. and Jana, Somnath and Karis, Olof and Freeland, John W. and Sarma, D. D.},
  journal = {Phys. Rev. B},
  volume = {97},
  issue = {16},
  pages = {165137},
  numpages = {7},
  year = {2018},
  month = {Apr},
  publisher = {American Physical Society},
  doi = {10.1103/PhysRevB.97.165137},
  url = {https://link.aps.org/doi/10.1103/PhysRevB.97.165137}
}

@article{PhysRevB.77.195104,
  title = {Effect of disorder on various physical properties of ${\text{Co}}_{2}\text{CrAl}$ Heusler alloy films: Experiment and theory},
  author = {Kudryavtsev, Y. V. and Uvarov, V. N. and Oksenenko, V. A. and Lee, Y. P. and Kim, J. B. and Hyun, Y. H. and Kim, K. W. and Rhee, J. Y. and Dubowik, J.},
  journal = {Phys. Rev. B},
  volume = {77},
  issue = {19},
  pages = {195104},
  numpages = {9},
  year = {2008},
  month = {May},
  publisher = {American Physical Society},
  doi = {10.1103/PhysRevB.77.195104},
  url = {https://link.aps.org/doi/10.1103/PhysRevB.77.195104}
}

@article{PhysRevB.107.184408,
  title = {Experimental observation of spin glass state in the highly disordered quaternary Heusler alloy FeRuMnGa},
  author = {Gupta, Shuvankar and Chakraborty, Sudip and Pakhira, Santanu and Biswas, Anis and Mudryk, Yaroslav and Kumar, Amit and Mukherjee, Bodhoday and Okram, Gunadhor S. and Das, Amitabh and Pecharsky, Vitalij K. and Mazumdar, Chandan},
  journal = {Phys. Rev. B},
  volume = {107},
  issue = {18},
  pages = {184408},
  numpages = {10},
  year = {2023},
  month = {May},
  publisher = {American Physical Society},
  doi = {10.1103/PhysRevB.107.184408},
  url = {https://link.aps.org/doi/10.1103/PhysRevB.107.184408}
}

@article{Kresse,
  title = {Efficient iterative schemes for \textit{ab initio} total-energy calculations using a plane-wave basis set},
  author = {G. Kresse and J. Furthm\"uller},
  journal = {Phys. Rev. B},
  volume = {54},
  issue = {16},
  pages = {11169-11186},
  year = {1996},
  month = {Oct},
  publisher = {American Physical Society},
  doi = {10.1103/PhysRevB.54.11169},
}

@article{PAW,
  title = {From ultrasoft pseudopotentials to the projector augmented-wave method},
  author = {G. Kresse and D. Joubert},
  journal = {Phys. Rev. B},
  volume = {59},
  issue = {3},
  pages = {1758--1775},
  year = {1999},
  month = {Jan},
  publisher = {American Physical Society},
  doi = {10.1103/PhysRevB.59.1758},
}

@article{PAWpotentials1,
  title = {Projector augmented-wave method},
  author = {P. E. Bl\"ochl},
  journal = {Phys. Rev. B},
  volume = {50},
  issue = {24},
  pages = {17953-17979},
  year = {1994},
  month = {Dec},
  publisher = {American Physical Society},
  doi = {10.1103/PhysRevB.50.17953},
}

@article{PBE,
  title = {Generalized Gradient Approximation Made Simple},
  author = {Perdew, John P. and Burke, Kieron and Ernzerhof, Matthias},
  journal = {Phys. Rev. Lett.},
  volume = {77},
  issue = {18},
  pages = {3865--3868},
  numpages = {0},
  year = {1996},
  month = {Oct},
  publisher = {American Physical Society},
  doi = {10.1103/PhysRevLett.77.3865},
  url = {https://link.aps.org/doi/10.1103/PhysRevLett.77.3865},
}

@article{wannier90,
title = {wannier90: A tool for obtaining maximally-localised {Wannier} functions },
journal = {Comput. Phys. Commun.},
volume = {178},
number = {9},
pages = {685 - 699},
year = {2008},
doi = {10.1016/j.cpc.2007.11.016},
author = {Arash A. Mostofi and Jonathan R. Yates and Young-Su Lee and Ivo Souza and David Vanderbilt and Nicola Marzari},
}

@article{tb2j,
title = {TB2J: A python package for computing magnetic interaction parameters},
journal = {Computer Physics Communications},
volume = {264},
pages = {107938},
year = {2021},
issn = {0010-4655},
doi = {https://doi.org/10.1016/j.cpc.2021.107938},
url = {https://www.sciencedirect.com/science/article/pii/S0010465521000679},
author = {Xu He and Nicole Helbig and Matthieu J. Verstraete and Eric Bousquet},
}

@article{vampire,
doi = {10.1088/0953-8984/26/10/103202},
url = {https://doi.org/10.1088/0953-8984/26/10/103202},
year = {2014},
month = {feb},
publisher = {IOP Publishing},
volume = {26},
number = {10},
pages = {103202},
author = {Evans, R F L and Fan, W J and Chureemart, P and Ostler, T A and Ellis, M O A and Chantrell, R W},
title = {Atomistic spin model simulations of magnetic nanomaterials},
journal = {Journal of Physics: Condensed Matter},}

@misc{SM,
  title        = {Supplementary Materials},
doi = {DOI and URL of Supplementary Materials will be provided by the publisher},
  url = {will be provided by publisher}
}

@article{AniruddhaDeb_2000,
doi = {10.1088/0953-8984/12/13/310},
url = {https://doi.org/10.1088/0953-8984/12/13/310},
year = {2000},
month = {apr},
publisher = {},
volume = {12},
number = {13},
pages = {2997},
author = {Aniruddha Deb and Yoshiharu Sakurai},
title = {Electronic 
structure of the Cu2MnAl 
Heusler alloy},
journal = {Journal of Physics: Condensed Matter},
}

@article{PhysRevB.108.054405,
  title = {$4d$ element induced improvement of structural disorder and development of weakly reentrant spin-glass behavior in NiRuMnSn},
  author = {Gupta, Shuvankar and Chakraborty, Sudip and Bhasin, Vidha and Pakhira, Santanu and Biswas, Anis and Mudryk, Yaroslav and Kumar, Amit and Barreteau, Celine and Crivello, Jean-Claude and Das, Amitabh and Jha, S. N. and Bhattacharyya, D. and Pecharsky, Vitalij K. and Alleno, Eric and Mazumdar, Chandan},
  journal = {Phys. Rev. B},
  volume = {108},
  issue = {5},
  pages = {054405},
  numpages = {11},
  year = {2023},
  month = {Aug},
  publisher = {American Physical Society},
  doi = {10.1103/PhysRevB.108.054405},
  url = {https://link.aps.org/doi/10.1103/PhysRevB.108.054405}
}

@article{PhysRevB.106.115148,
  title = {Coexisting structural disorder and robust spin-polarization in half-metallic FeMnVAl},
  author = {Gupta, Shuvankar and Chakraborty, Sudip and Pakhira, Santanu and Barreteau, Celine and Crivello, Jean-Claude and Bandyopadhyay, Bilwadal and Greneche, Jean Marc and Alleno, Eric and Mazumdar, Chandan},
  journal = {Phys. Rev. B},
  volume = {106},
  issue = {11},
  pages = {115148},
  numpages = {12},
  year = {2022},
  month = {Sep},
  publisher = {American Physical Society},
  doi = {10.1103/PhysRevB.106.115148},
  url = {https://link.aps.org/doi/10.1103/PhysRevB.106.115148}
}

@article{PhysRevB.108.054430,
  title = {Origin of magnetic ordering in half-Heusler RuMnGa},
  author = {Chakraborty, Sudip and Gupta, Shuvankar and Pakhira, Santanu and Biswas, Anis and Mudryk, Yaroslav and Choudhary, Renu and Kumar, Amit and Das, Amitabh and Mazumdar, Chandan},
  journal = {Phys. Rev. B},
  volume = {108},
  issue = {5},
  pages = {054430},
  numpages = {12},
  year = {2023},
  month = {Aug},
  publisher = {American Physical Society},
  doi = {10.1103/PhysRevB.108.054430},
  url = {https://link.aps.org/doi/10.1103/PhysRevB.108.054430}
}

@article{PhysRevB.108.245151,
  title = {Observation of charge transfer induced large enhancement of magnetic moment in a structurally disordered inverse Heusler alloy ${\mathrm{Fe}}_{2}\mathrm{RuGe}$},
  author = {Chakraborty, Sudip and Gupta, Shuvankar and Bhasin, Vidha and Pakhira, Santanu and Barreteau, C\'eline and Crivello, Jean-Claude and Jha, Shambhu Nath and Bhattacharyya, Dibyendu and Avdeev, Maxim and Paul-Boncour, Val\'erie and Greneche, Jean-Marc and Alleno, Eric and Mazumdar, Chandan},
  journal = {Phys. Rev. B},
  volume = {108},
  issue = {24},
  pages = {245151},
  numpages = {13},
  year = {2023},
  month = {Dec},
  publisher = {American Physical Society},
  doi = {10.1103/PhysRevB.108.245151},
  url = {https://link.aps.org/doi/10.1103/PhysRevB.108.245151}
}

@article{Patra_2026,
doi = {10.1088/1361-648X/ae60bb},
url = {https://doi.org/10.1088/1361-648X/ae60bb},
year = {2026},
month = {apr},
publisher = {IOP Publishing},
volume = {38},
number = {16},
pages = {165807},
author = {Patra, Arka and Dan, Shovan and Modak, Mantu and Kaczorowski, D and Mazumdar, Chandan},
title = {Origin of low magnetic moment in quaternary Heusler alloy NiFeTiGa},
journal = {Journal of Physics: Condensed Matter},
}

@article{PhysRevB.98.184419,
  title = {Structural correlation with the Griffiths phase in disordered magnetic systems},
  author = {Ghosh, Krishanu and Mazumdar, Chandan and Ranganathan, R. and Mukherjee, S. and De Raychaudhury, Molly},
  journal = {Phys. Rev. B},
  volume = {98},
  issue = {18},
  pages = {184419},
  numpages = {8},
  year = {2018},
  month = {Nov},
  publisher = {American Physical Society},
  doi = {10.1103/PhysRevB.98.184419},
  url = {https://link.aps.org/doi/10.1103/PhysRevB.98.184419}
}

\end{document}


\title{Supplementary Materials for\\
Enhancement of local ordering temperature due to non-magnetic disorder in anti-Heusler alloy Al$_2$MnCu}
\author{Soumya Bhowmik}
\email{soumyabhowmikwb@gmail.com}
\affiliation{Condensed Matter Physics Division, Saha Institute of Nuclear Physics,
              A CI of Homi Bhabha National Institute, 1/AF Bidhannagar, Kolkata 700064, India}
\author{Suvayan Saha}
\affiliation{Condensed Matter Physics Division, Saha Institute of Nuclear Physics,
              A CI of Homi Bhabha National Institute, 1/AF Bidhannagar, Kolkata 700064, India}
\author{Saikat Debnath}
\affiliation{Department of Physics, M.V. College, Buxar, Bihar 802101, India}
\author{Arka Patra}
\affiliation{Condensed Matter Physics Division, Saha Institute of Nuclear Physics,
              A CI of Homi Bhabha National Institute, 1/AF Bidhannagar, Kolkata 700064, India}
\author{Biswarup Satpati}
\affiliation{Condensed Matter Physics Division, Saha Institute of Nuclear Physics,
              A CI of Homi Bhabha National Institute, 1/AF Bidhannagar, Kolkata 700064, India}
\author{Maxim Avdeev}
\affiliation{Australian Nuclear Science and Technology Organisation, Locked Bag 2001, Kirrawee DC, New South Wales 2232, Australia}
\affiliation{School of Chemistry, The University of Sydney, Sydney, New South Wales 2006, Australia}
\author{Shishir Kumar Pandey}
\email{shishir.kr.pandey@gmail.com}
\affiliation{Department of General Sciences (Physics), Birla Institute of Technology and Science, Pilani, Dubai Campus, Dubai International Academic City, Dubai 345055, United Arab Emirates}
\affiliation{Department of Physics, Birla Institute of Technology and Science, Pilani, Hyderabad Campus, Jawahar Nagar, Kapra Mandal, Medchal District, Telangana 500078, India}
\author{Chandan Mazumdar}
\email{chandan.mazumdar@saha.ac.in}
\affiliation{Condensed Matter Physics Division, Saha Institute of Nuclear Physics,
              A CI of Homi Bhabha National Institute, 1/AF Bidhannagar, Kolkata 700064, India}

\date{\today}
\maketitle
~\\[1em]

\section{Critical analysis}
\begin{figure}
  \centering
  \includegraphics[width=\textwidth]{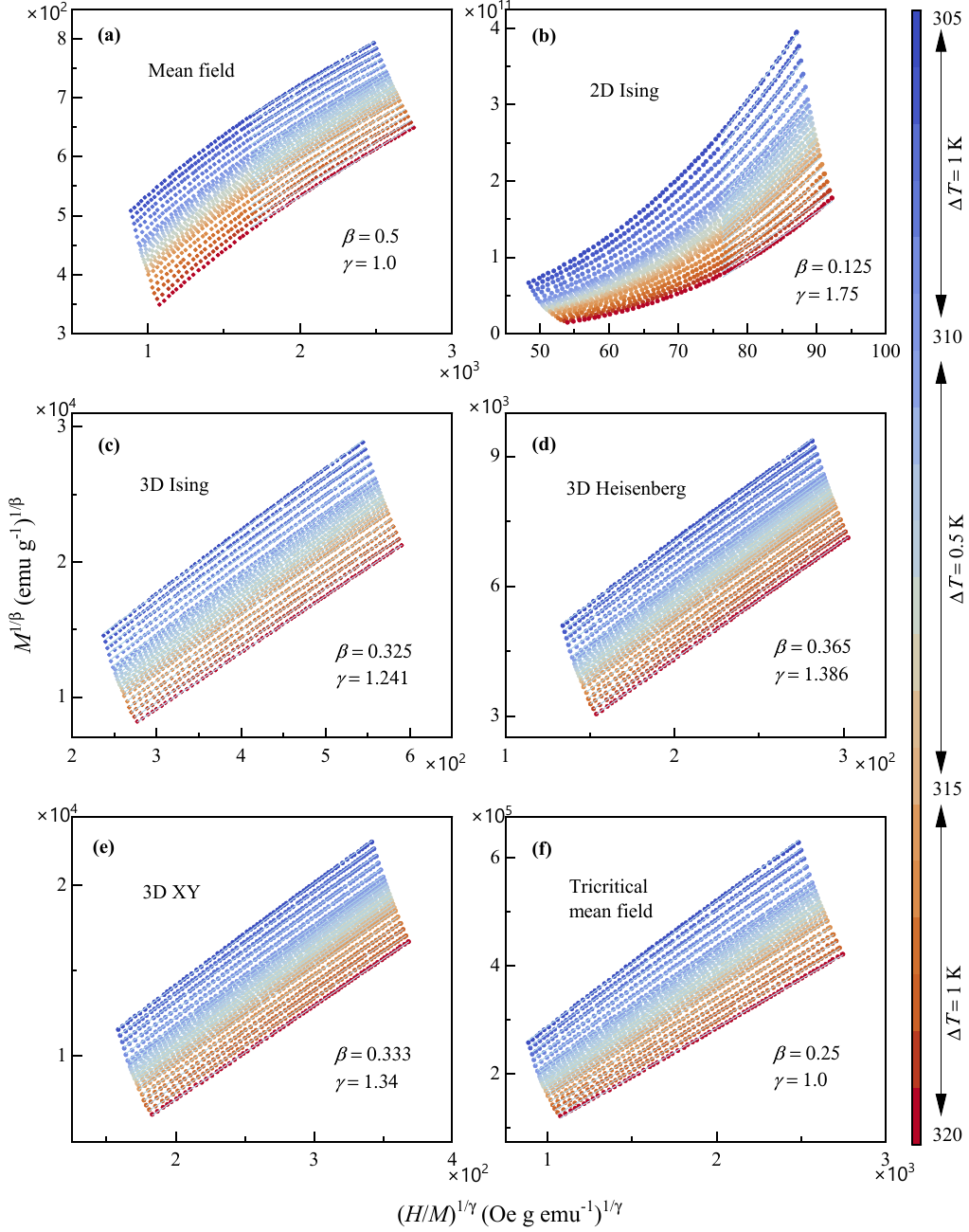}
  \caption{Modified Arrott plot for different universal classes, i.e, (a) Mean field theory ($\beta$= 0.5, $\gamma$= 1), (b) 2D Ising model ($\beta$= 0.125, $\gamma$= 1.75), (c) 3D Ising model ($\beta$= 0.325, $\gamma$= 1.241), (d) 3D Heisenberg model ($\beta$= 0.365, $\gamma$= 1.386), (e) 3D XY model ($\beta$= 0.333, $\gamma$= 1.34), (f) Tricritical mean-field theory ($\beta$= 0.25, $\gamma$= 1). Blue lines are the linear fit.}\label{All Map}
\end{figure}

\begin{figure}
  \centering
  \includegraphics[width=0.5\textwidth]{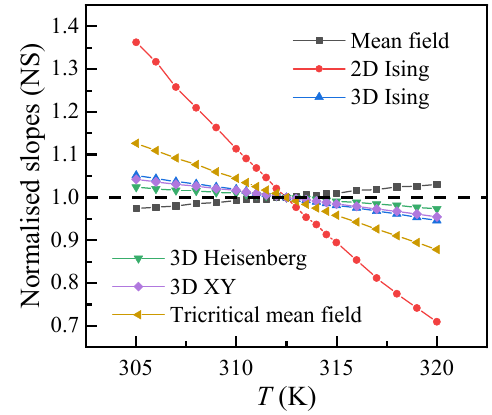}
  \caption{Normalised slope of different universal classes.}\label{Normalised slope}
\end{figure}

\begin{figure}
  \centering
  \includegraphics[width=0.5\textwidth]{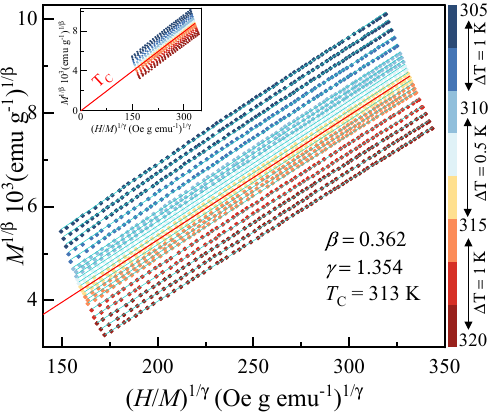}
  \caption{The modified Arrot plot of Al$_2$MnCu, constructed by $\beta$= 0.362, $\gamma$= 1.354. The inset shows that the critical isotherm passes through the origin.}\label{Final MAP}
\end{figure}

To start with the right choice of initial critical exponents, the modified Arrott plots (MAP) are constructed employing several universal classes of models which are Mean field ($\beta$= 0.5, $\gamma$= 1), 2D Ising ($\beta$= 0.125, $\gamma$= 1.75), 3D Ising ($\beta$= 0.325, $\gamma$= 1.241), 3D Heisenberg ($\beta$= 0.365, $\gamma$= 1.386), 3D XY ($\beta$= 0.333, $\gamma$= 1.34), and tricritical mean-field ($\beta$= 0.25, $\gamma$= 1) model [Fig.~\ref{All Map}]~\cite{PhysRevB.105.214407}. The best choice of initial critical exponents, for which the MAP becomes a set of parallel straight lines, i.e, the normalized slope, $NS [= S(T)/S(T_{\rm C})$], will follow the ideal value 1 with the least deviation~\cite{PhysRevB.99.214414}. By examining the $NS(T)$ for different universal models[Fig.~\ref{Normalised slope}], the initial critical exponent has been chosen as $\beta$= 0.365, $\gamma$= 1.386 (3D Heisenberg model). Taking these initial parameters, after multiple iterations of simultaneous fitting with the modified Arrott plot and power law, the critical parameter converges to $\beta$= 0.362, $\gamma$= 1.354 [Fig.~\ref{Final MAP}], where the critical isotherm passes through the origin [Fig.~\ref{Final MAP}(inset)]. To find out the dimensionality and range of the interactions, renormalization-group theory was applied, and the estimated parameters, obtained from Eq.~10 of the main manuscript, have been tabulated in Table~\ref{tab:critical}.

\begin{table}[b]
\centering
\renewcommand{\arraystretch}{1.15}
\begin{tabular}{cccccc}
\hline \hline
\multicolumn{6}{l}{Space group $Fm\bar{3}m$, No. 225} \\
\multicolumn{6}{l}{Lattice parameter $a=b=c=$ 5.947(7) \AA} \\
\hline

Symmetry site
& Wyckoff position
& $(x,y,z)$
& Element
& \begin{tabular}[c]{@{}c@{}}Occupancy\\(fraction)\end{tabular}
& \begin{tabular}[c]{@{}c@{}}Occupancy\\(\%)\end{tabular} \\
\hline

\multirow{6}{*}{Octahedral}
& \multirow{3}{*}{$4a$}
& \multirow{3}{*}{$(0,0,0)$}
& Mn & 0.500(1) & 50 \\
& & & Cu & 0.453(3) & 45.3 \\
& & & Al & 0.047(3) & 4.7 \\

\cline{2-6}

& \multirow{3}{*}{$4b$}
& \multirow{3}{*}{$(0.5,0.5,0.5)$}
& Mn & 0.500(1) & 50 \\
& & & Cu & 0.453(3) & 45.3 \\
& & & Al & 0.047(3) & 4.7 \\

\hline

\multirow{2}{*}{Tetrahedral}
& \multirow{2}{*}{$8c$}
& \multirow{2}{*}{$(0.25,0.25,0.25)$}
& Al & 0.953(4) & 95.3 \\
& & & Cu & 0.047(2) & 4.7 \\

\hline \hline
\end{tabular}
\caption{Structural parameters obtained from Rietveld refinement.}
\label{tab:structure}
\end{table}

\begin{table*}
\centering
\begin{tabular*}{\textwidth}{@{\hspace{5pt}}@{\extracolsep{\fill}}cccccccc@{\extracolsep{\fill}}@{\hspace{5pt}}}
\hline
$d$ & $n$ & Model/compound & $\beta$ & $\sigma$
& $\nu\,(=\frac{\gamma}{\sigma})$ & $\frac{\beta}{\nu}$ & $\frac{\gamma}{\nu}$ \\
\hline

\multirow{5}{*}{3}
& 1 & \multirow{4}{*}{Al$_2$MnCu}
& 0.318 & 2.04 & 0.663 & 0.479 & 2.039 \\

& 2 & & 0.360 & 1.957 & 0.691 & 0.521 & 1.956 \\

& 3 & & 0.387 & 1.908 & 0.709 & 0.546 & 1.907 \\

& $\infty$ & & 0.497 & 1.729 & 0.782 & 0.636 & 2.211 \\

& & Mean field & 0.5 & 2.000 & 0.5 & 1 & 2 \\

2 & 1 & 2D Ising & 0.125 & 1.750 & 1 & 0.125 & 1.750 \\

3 & 1 & 3D Ising & 0.325 & 1.970 & 0.63 & 0.516 & 1.969 \\

3 & 3 & 3D Heisenberg & 0.365 & 1.966 & 0.705 & 0.518 & 1.966 \\

3 & 3 & 3D XY & 0.333 & 2 & 0.67 & 0.497 & 2.000 \\

& & Tricritical mean field & 0.25 & 2 & 0.500 & 0.500 & 2 \\

\hline
\end{tabular*}
\caption{Critical exponents and related parameters for different models and compounds.}
\label{tab:critical}
\end{table*}

\section{DC Magnetization}
\begin{figure}
  \centering
  \includegraphics[width=0.5\textwidth]{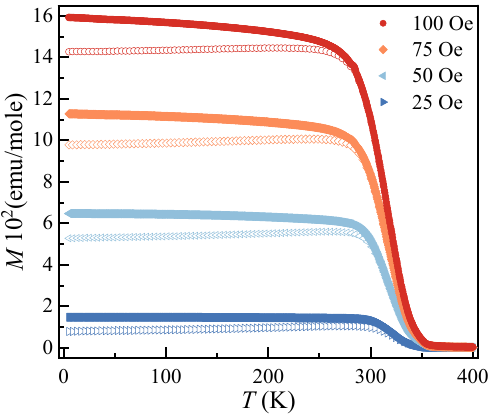}
  \caption{Temperature dependent magnetization, $M(T)$ curve for different low applied field.}\label{MT}
\end{figure}

\begin{figure}
  \centering
  \includegraphics[width=0.5\textwidth]{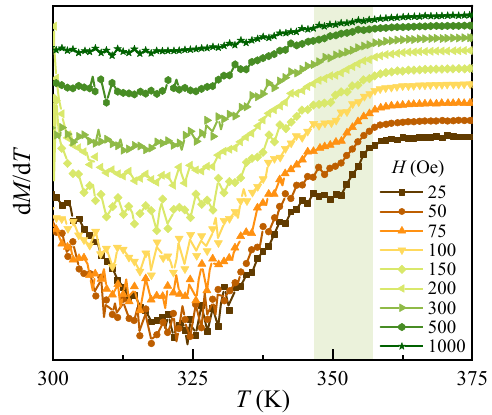}
  \caption{d$M$/d$T$ curve for various applied fields. The curves are translated vertically with reference to the 1000~Oe curve, for better visualisation. The shaded region is the temperature regime where the secondary magnetic phase appears. In this region, the additional kink observed for low applied field and vanishes gradually as the field increases.}\label{DChi_dT}
\end{figure}

\begin{figure}
  \centering
  \includegraphics[width=\textwidth]{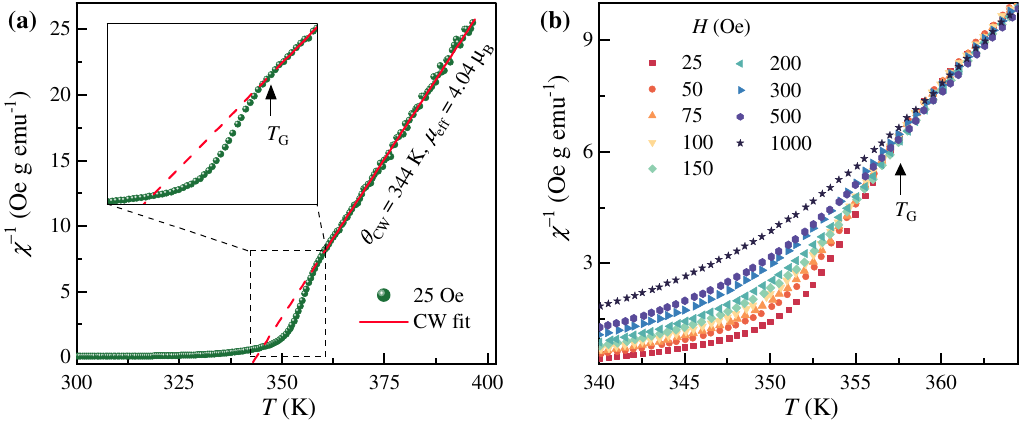}
  \caption{(a) The Curie-Weiss fit of $\chi'^{-1}(T)$ at 25~Oe, in the high-temperature region ($T>$360~K). The inset shows the depression of the curve from linear CW behaviour. (b) The gradual suppression of the downturn under a higher applied field.}\label{GP}
\end{figure}

The $M(T)$ data of Al$_2$MnCu has been measured under different applied field [Fig.~\ref{MT}]. The first-order derivative of $M$ with respect to $T$, (d$M$/d$T$), reveals, along with a dip at $T_{\rm C}$, an additional secondary kink above $T_{\rm C}$, at $T_{\rm P}\sim$344~K is present, which is suppressed under the application of higher field [Fig.~\ref{DChi_dT}]. This additional peak is indicating the existence of a secondary magnetic phase in this ferromagnetic compound. The presence of this secondary magnetic phase is further substantiated by the inverse susceptibility plot, where the high-temperature region is well described by the Curie-Weiss law [$\chi^{-1}(T)= C/(T-\theta_{CW})$] of a paramagnet, whereas below a certain temperature $T_{\rm G}\sim$358~K, the curve exhibits a downturn from linearity [Fig.~\ref{GP}(a)]. The non-linear downturn, above $T_{\rm C}$, is suppressed under a higher amplitude of the applied field, which is a fingerprint of the Griffiths phase [Fig.~\ref{GP}(b)]~\cite{PhysRevB.98.184419}.

\begin{figure}
  \centering
  \includegraphics[width=0.5\textwidth]{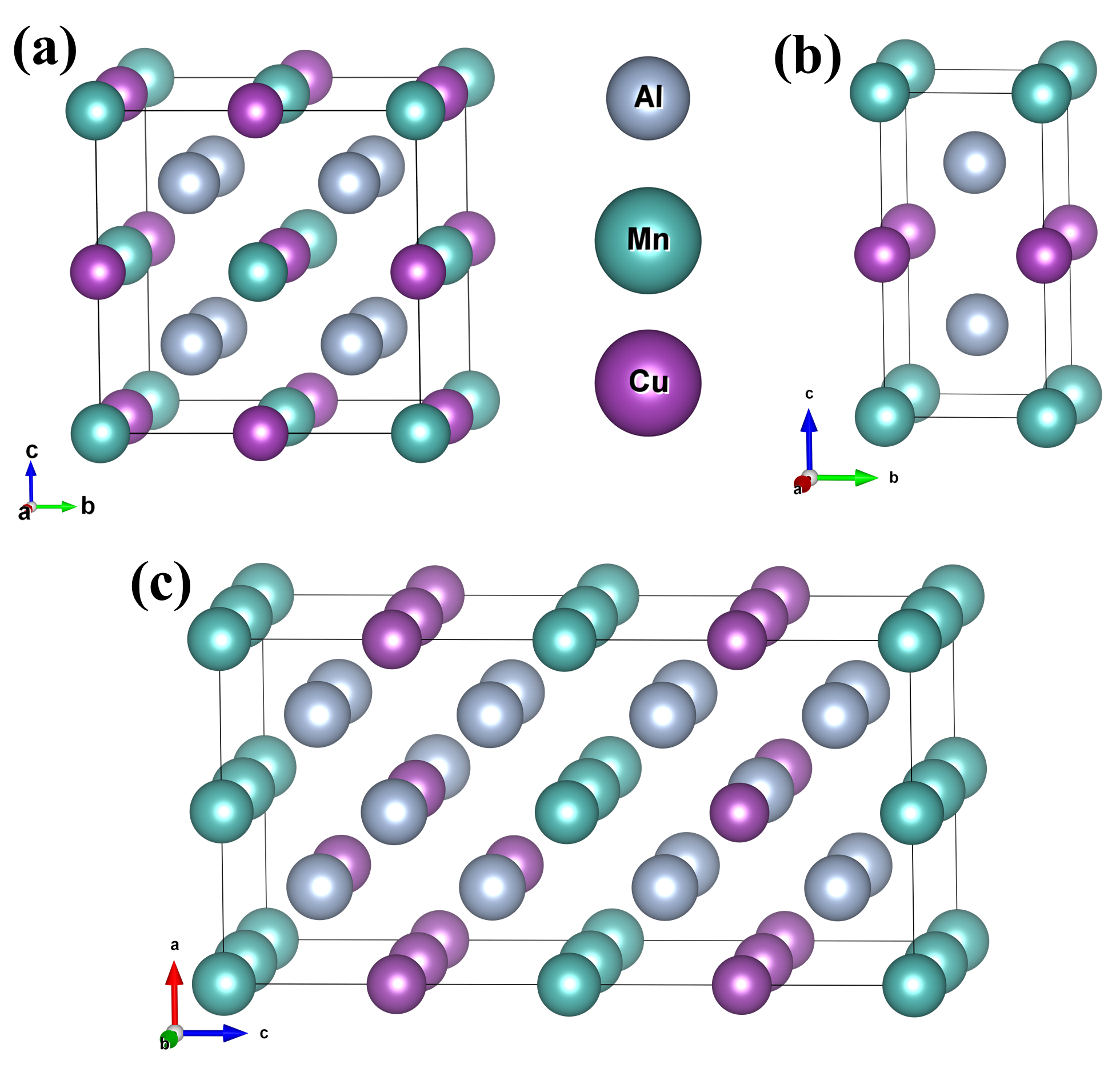}
  \caption{Simulated (a) L2$_1$ ordered, (b) B2-type disordered, and (c) B2+12.5\% Cu-Al disorder structures for theoretical calculations.}\label{TS}
\end{figure}
\section{ac susceptibility}
\begin{figure}
  \centering
  \includegraphics[width=\textwidth]{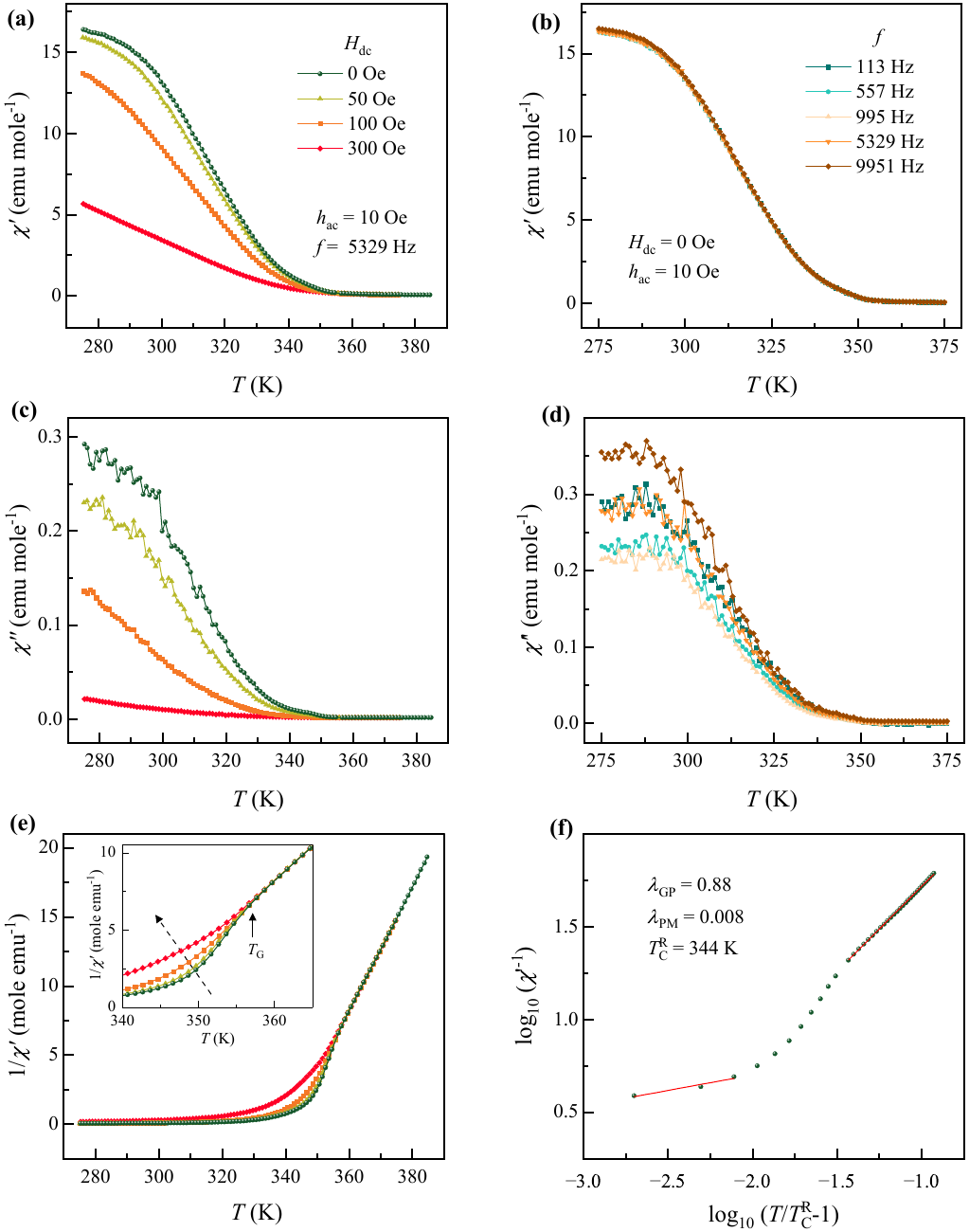}
  \caption{(a) Real ($\chi'(T)$) and (c) imaginary ($\chi''(T)$) part of ac susceptibility data measured under various applied dc field. (b) $\chi'(T)$ and (d) $\chi''(T)$ data for different frequency. (e) The down-turn in $\chi'^{-1}(T)$ plot gradually diminishes under higher applied field (inset). (f) The power law fitting of 25~Oe $\chi'^{-1}(T)$ curve.}\label{AC}
\end{figure}

The frequency ($f$) and dc field ($H_{dc}$) dependent real ($\chi'(T)$) and imaginary ($\chi''(T)$) part of ac susceptibility is presented here. Fig.~\ref{AC}(b) and (d) reveals no frequency dependence in the $T_{\rm C}\leq T \leq T_{\rm G}$ region, rules out the glassy or super-paramagetic phase. The $H_{dc}$-dependent study unveiled an anomaly at $T_{\rm C}\leq T \leq T_{\rm G}$ regime [Fig.~\ref{AC}(a) and (c)], more prominently observed in  $\chi'^{-1}(T)$ plot [Fig.~\ref{AC}(e)], which also shows a non-linear downturn, resembling the dc magnetization. Further, the power law fitting [$\chi^{-1}(T) \propto (T-T_{\rm C}^{\rm R})^{1-\lambda}$, see main manuscript] yielded $\lambda_{GP}\sim$0.88, establish the Griffiths phase [Fig.~\ref{AC}(f)] in the anti-Heusler alloy Al$_2$MnCu.

\FloatBarrier
\bibliographystyle{apsrev4-2}
\bibliography{SM_A2MC}